\documentclass[12pt]{article}
\usepackage{epsfig}           
\usepackage{amsfonts}
\usepackage{amssymb}

\begin{document}
\begin{flushright}  {~} \\[-1cm]
{\sf hep-th/0002195}\\
{\sf ETH-TH/00-1}\\
{\sf February 2000} \end{flushright}

\begin{center} \vskip 15mm
{\Large\bf New applications of the chiral anomaly
  \footnote{This review is
     dedicated to the memory of Louis Michel, the theoretician and the
     friend.}}\\[22mm]
{\large J\"urg Fr\"ohlich and Bill Pedrini}\\[7mm]
Institut f\"ur Theoretische Physik \\
ETH H\"onggerberg\\ CH\,--\,8093\, Z\"urich\\[5mm]
{\it E-mail: juerg@itp.phys.ethz.ch; ~pedrini@itp.phys.ethz.ch}
\end{center}
\vskip 15mm

\begin{abstract}
We describe consequences of the chiral anomaly in the
  theory of quantum wires, the (quantum) Hall effect, and of a
  four-dimensional cousin of the Hall effect. We explain which aspects 
  of conductance quantization are related to the chiral anomaly. The
  four-dimensional analogue of the Hall effect involves the axion
  field, whose time derivative can be interpreted as a (space-time
  dependent) difference of chemical potentials of left-handed and
  right-handed charged fermions. Our four-dimensional analogue of the
  Hall effect may play a significant r\^ole in explaining the origin
  of large magnetic fields in the (early) universe.
\end{abstract}
\vskip 26mm

\section{What is the chiral anomaly?}

The chiral abelian anomaly has been discovered, in the past century,
by Adler, Bell and Jackiw, after earlier work on $\pi^0$-decay
starting with Steinberger and Schwinger; see e.g. [1] and references given
there. 
It has been rederived in many different ways of varying degree of
mathematical rigor by many people. Diverse physical implications,
especially in particle physics, have been discussed. It is hard to
imagine that one may still be able to find new, interesting
implications of the chiral anomaly that specialists have not been
aware of, for many years. Yet, until very recently --- in the past
century, but only two to three years ago --- this turned out to be
possible, and we suspect that further applications may turn up in the
future! This little review is devoted to a discussion of physical
implications of the chiral anomaly that have been discovered
recently.

Before we turn to physics, we recall what is meant by ``chiral
(abelian) anomaly''. 
In general terms, one speaks of an {\it anomaly} if some quantum
theory violates a symmetry present at the classical level, (i.e., in
the limit where $\hbar\to 0$). By ``{\it violating a symmetry}'' one means 
that it is impossible to construct a unitary representation of the
symmetry transformations of the classical system underlying some
quantum theory on the Hilbert space of pure state vectors of the
quantum theory. (By ``violating a {\it dynamical} symmetry'' is meant
that it is impossible to construct such a representation that commutes 
with the {\it unitary time evolution} of the quantum theory.)

It is quite clear that understanding anomalies can be viewed as a
problem in group cohomology theory. A fundamental example of an
anomalous symmetry group is the group of all symplectic
transformations of the phase space of a classical Hamiltonian system
underlying some quantum theory.

The anomalies considered in this review are ones connected with
infinite-dimensional groups of gauge transformations which are
symmetries of some classical Lagrangian systems with infinitely many
degrees of freedom (Lagrangian field theories). Thus, we consider a
theory of charged, massless fermions coupled to an external
electromagnetic field in Minkowski space-time of even dimension
$2n$. Let $\gamma^0, \gamma^1, \ldots, \gamma^{2n-1}$ denote the usual 
Dirac matrices, and define
\begin{equation}
\gamma\;:=\;i\,\gamma^0\,\gamma^1\ldots\gamma^{2n-1} \ .
\label{1.1}
\end{equation}
Then $\gamma$ anti-commutes with the {\it covariant Dirac operator}
\begin{equation}
D\;:=\; i\,\gamma^\mu\,\left( \partial_\mu - i A_\mu\right) \ ,
\label{1.2}
\end{equation}
where $A$ is the vector potential of the external electromagnetic
field. Let $\psi(x)$ denote the Dirac spinor field and $\bar{\psi}(x)$ 
the conjugate field. We define the vector current, ${\mathcal J}^\mu$, 
and the axial vector current $\widetilde{{\mathcal J}}^\mu$, by
\begin{equation}
{\mathcal J}^\mu\;:=\;\bar{\psi} \gamma^\mu \psi \ , \
  \widetilde{{\mathcal J}}^\mu \;:=\; \bar{\psi} \gamma^\mu \gamma
    \psi \ .
\label{1.3}
\end{equation}
At the classical level, these currents are {\it conserved},
\begin{equation}
\partial_\mu {\mathcal J}^\mu\;=\; 0~,~~\partial_\mu
\widetilde{{\mathcal J}}^\mu\;=\; 0~,
\label{1.4}
\end{equation}
on solutions of the equations of motion, $(D \psi = 0)$. The
conservation of the vector current is intimately connected with the
electromagnetic {\it gauge invariance} of the theory,
\begin{eqnarray}
&&{} \psi (x) \mapsto e^{i\chi(x)} \psi (x)~,~~\bar{\psi}(x) \mapsto
\bar{\psi}(x)\; e^{-i\chi(x)} \nonumber \\
&&{}~~~~~~A_\mu(x) \mapsto A_\mu (x) + \partial_\mu \chi (x)~, 
\label{1.5}
\end{eqnarray}
where $\chi(x)$ is a test function on space-time. When $\chi$ is
constant in $x$ the transformations (\ref{1.5}) are a global symmetry
of the classical theory
corresponding to the conserved quantity
\begin{equation}
Q\;=\;\int d\underline{x}\; {\mathcal J}^0
\left(x^0,\underline{x}\right) 
\label{1.6}
\end{equation}
which is the {\it electric charge}. The conservation of $Q$
(independence of $x^0$) follows, of course, from the fact that the
Noether current ${\mathcal J}^\mu$ associated with (\ref{1.5}) satisfies the
continuity equation (\ref{1.4}). 

The conservation of the axial vector current $\widetilde{{\mathcal
  J}}^\mu$, in the classical theory, is connected with the invariance
of the theory under {\it local chiral rotations}
\begin{eqnarray}
&&{} \psi (x) \mapsto e^{i\alpha(x)\gamma} \psi (x),~~ \bar{\psi}(x)
\mapsto \bar{\psi} (x)\;e^{i\alpha(x)\gamma} \nonumber\\
&&{}~~~~~~~~~~~A_\mu (x) \mapsto A_\mu (x) + \gamma \partial_\mu
  \alpha (x)~,
\label{1.7}
\end{eqnarray}
where $\alpha (x)$ is a test function on space-time. In particular,
when $\alpha$ is a constant the transformations (\ref{1.7}) are a
global symmetry of the classical theory corresponding to the conserved
charge  
\begin{equation}
\widetilde{Q}\;=\;\int d\underline{x}\; \widetilde{{\mathcal J}}^0
\left( x^0,\underline{x}\right)
\label{1.8}
\end{equation}
(which, according to (\ref{1.4}), is independent of $x^0$).

It turns out that, in the quantum theory, the local chiral rotations
(\ref{1.7})  do {\it not} leave quantum-mechanical transition
amplitudes invariant, and the axial vector current
$\widetilde{{\mathcal J}}^\mu$ is {\it not} a conserved current, for
arbitrary external electromagnetic fields. This phenomenon is called
{\it chiral} (abelian) {\it anomaly}.

Let us see where the chiral anomaly comes from, for theories in two and 
four space-time dimensions. We start with the discussion of {\it
  two-dimensional theories}. We consider a quantum theory which has a
conserved vector current ${\mathcal J}^\mu$ and --- {\it if the
  external electromagnetic field vanishes} --- a conserved axial
vector current $\widetilde{{\mathcal J}}^\mu$, i.e.,
\begin{equation}
\partial_\mu \,{\mathcal J}^\mu \;=\;0~,~~\partial_\mu\,
\widetilde{{\mathcal J}}^\mu\;=\;0~.
\label{1.9}
\end{equation}

In two space-time dimensions, ${\mathcal J}^\mu$ and
$\widetilde{{\mathcal J}}^\mu$ are related to each other by
\begin{equation}
\widetilde{{\mathcal J}}^\mu \;=\; \varepsilon^{\mu\nu}\, {\mathcal
J}_\nu 
\label{1.10}
\end{equation}
where $\varepsilon^{00} = \varepsilon^{11} = 0~,~~
\varepsilon^{01}=-\varepsilon^{10}=1$. The continuity equation
\[
\partial_\mu\,{\mathcal J}^\mu\;=\;0
\]
has the general solution
\begin{equation}
{\mathcal J}^\mu (x)\;=\;\frac{q}{2\pi}~\varepsilon^{\mu\nu}
\left(\partial_\nu\,\varphi\right) (x)~,
\label{1.11}
\end{equation}
where $\varphi (x)$ is an arbitrary scalar field on 
space-time, and $q$ denotes the electric charge. Using
eqs.~(\ref{1.11})  and (\ref{1.10}) and the continuity equation, 
\[
\partial_\mu\,\widetilde{{\mathcal J}}^\mu\;=\;0~,
\]
for the axial vector current, we find that the field $\varphi$ must
obey the equation of motion
\begin{equation}
\square ~\varphi (x)\;=\;0~.
\label{1.12}
\end{equation}
Thus, if the vector- and axial vector currents are conserved then the
potential $\varphi$ of the vector current is a {\it massless free
  field}. The theory of the massless free field is an example of a
Lagrangian field theory. It has an action functional, $S$, given by
\begin{equation}
S (\varphi)\;=\; \frac{1}{4\pi} \int d^2x\;\left(
  \partial_\mu\varphi\right) (x) \left(\partial^\mu \varphi\right)
(x)~. 
\label{1.13}
\end{equation}
The ``momentum'', $\pi (x)$, canonically conjugate to $\varphi (x)$ is 
defined, as usual, by
\begin{equation}
\pi (x) \;=\; \delta S (\varphi)\big/ \delta \left( \partial_0 \varphi 
  (x)\right) \;=\; \frac{1}{2\pi}~ \frac{\partial \varphi
  (x)}{\partial t}~,
\label{1.14}
\end{equation}
where $t=x^0$ denotes time; (the ``velocity of light'' $c=1$). In the
quantum theory, $\varphi$ and $\pi$ are operator-valued distributions
on Fock space satisfying the equal-time {\it canonical commutation
  relations}
\begin{equation}
\left[ \pi \left( t, \underline{x}\right)~,~~\varphi \left( t,
    \underline{y}\right)\right] \;=\; - i \delta \left( \underline{x}
  - \underline{y}\right)~.
\label{1.15}
\end{equation}
Since
\[
{\mathcal J}^\mu (x) \;=\; \frac{q}{2\pi}~\varepsilon^{\mu\nu} \left(
  \partial_\nu \varphi\right) (x)~,
\]
and
\[
\widetilde{{\mathcal J}}^\mu (x) \;=\; \varepsilon^{\mu\nu} {\mathcal
  J}_\nu (x) \;=\; \frac{q}{2\pi}~ \left( \partial^\mu \varphi\right)
(x)~, 
\]
eq.~(\ref{1.15}) yields the well known {\it anomalous commutator}
\begin{equation}
\left[ {\mathcal J}^0 \left( t,\underline{x}\right)~,
  ~~\widetilde{{\mathcal J}}^0 \left( t, \underline{y}\right)\right]
\;=\; i~\frac{q^2}{2\pi}~\delta' \left( \underline{x}
-\underline{y}\right)~. 
\label{1.16}
\end{equation}

Next, we imagine that the system is coupled to a classical external
electric field $E(x)$. In two space-time dimensions, the electric
field is given in terms of the electromagnetic vector potential
$A_\mu$ by
\begin{equation}
E(x) \;=\; \varepsilon^{\mu\nu} \left( \partial_\mu A_\nu\right)
(x)~. 
\label{1.17}
\end{equation}
The action functional for the theory in an external electric field is
given by
\begin{eqnarray}
S \left( \varphi, A\right)
&=& \frac{1}{4\pi} \int d^2x \left( \partial_\mu \varphi\right) (x)
\left(\partial^\mu\varphi\right) (x)~+~\frac 1 q \int d^2x {\mathcal
  J}^\mu (x) A_\mu (x) \nonumber\\
&=& \frac{1}{4\pi} \int d^2x \left\{ \left( \partial_\mu\varphi\right) 
  (x) \left(\partial^\mu\varphi\right)(x)~+~2\, \varepsilon^{\mu\nu}
  \partial_\nu \varphi (x) A_\mu (x)\right\}~.
\label{1.18}
\end{eqnarray}
The equation of motion (Euler-Lagrange equation) obtained from the
action function (\ref{1.18}) is
\begin{equation}
\square~\varphi (x) \;=\; E (x)~.
\label{1.19}
\end{equation}
Using (\ref{1.10}) and (\ref{1.11}), we see that equation~(\ref{1.19}) 
is 
equivalent to 
\begin{equation}
\partial_\mu \widetilde{{\mathcal J}}^\mu (x) \;=\; \frac{q}{2\pi}~E
(x)~, 
\label{1.20}
\end{equation}
i.e., the axial vector current {\it fails} to be conserved in a
non-vanishing external electric field $E$. Equation (\ref{1.20}) is
the standard expression of the chiral anomaly in two space-time
dimensions.

From the currents ${\mathcal J}^\mu$ and $\widetilde{{\mathcal
    J}}^\mu$ one can construct chiral currents, ${\mathcal
  J}_\ell^\mu$ and ${\mathcal J}_r^\mu$, for left-moving and
right-moving modes by setting
\begin{equation}
{\mathcal J}_\ell^\mu \;=\; {\mathcal J}^\mu -
  \widetilde{{\mathcal J}}^\mu~,~~ {\mathcal J}_r^\mu \;=\;
{\mathcal J}^\mu + \widetilde{{\mathcal
      J}}^\mu~. 
\label{1.21}
\end{equation}
They satisfy the equations
\begin{equation}
\partial_\mu {\mathcal J}_{\ell/r}^\mu \;=\; \mp~\frac{q}{2\pi}~E(x)~. 
\label{1.22}
\end{equation}

From eqs.~(\ref{1.17}) and (\ref{1.22}) we infer that one can define
modified chiral currents, $\widehat{{\mathcal J}}_{\ell/r}^\mu$, which 
{\it are} conserved: 
\begin{equation}
\widehat{{\mathcal J}}_{\ell/r}^\mu (x) \;:=\; {\mathcal
  J}_{\ell/r}^\mu \;\pm\; \frac{q}{2\pi}~ \varepsilon^{\mu\nu} A_\nu
(x)~. 
\label{1.23}
\end{equation}
Then 
\[
\partial_\mu \widehat{{\mathcal J}}_{\ell/r}^\mu (x) \;=\; 0~,
\]
but $\widehat{{\mathcal J}}_{\ell/r}^\mu$ fail to be
gauge-invariant. Nevertheless the conserved charges, 
\begin{equation}
N_\ell \;:= \int d\underline{x} \;\widehat{{\mathcal J}}_\ell^0 \left(
  t, \underline{x}\right)~, ~~N_r\;:= \int d\underline{x}\;
\widehat{{\mathcal J}}_r^0 \left( t, \underline{x}\right)~, 
\label{1.24}
\end{equation}
{\it are} gauge-invariant. They count the total electric charge of
left-moving and of right-moving modes, respectively, present in a
physical state of the system. 

The anomalous commutators are given by 
\[
\left[ {\mathcal J}^0 \left( t, \underline{x}\right)~,
  ~~\widehat{{\mathcal J}}_{\ell /r}^0 \left( t,
    \underline{y}\right)\right] \;=\; \mp\; i~\frac{q^2}{2\pi}~\delta'
\left( \underline{x} - \underline{y}\right)~.
\]

The left-moving / right-moving {\it charged fields} of the theory can
be expressed as normal-ordered exponentials of spatial integrals of
$\frac 1 q~\widehat{{\mathcal J}}_{\ell/r}^0 (x)$, i.e., as vertex
operators; they transform correctly under gauge transformations.

This completes our review of the chiral anomaly and of anomalous
commutators in two dimensions, and we now turn to {\it four}- (or
higher-) {\it dimensional systems}. 

We consider charged, massless Dirac fermions described by a Dirac
spinor field $\psi (x)$ and its conjugate field $\bar{\psi} (x) =
\psi^* (x) \gamma_0$. We study the effect of coupling these fields to external vector- and axial-vector potentials, $A_\mu$ and $Z_\mu$,
respectively. The theory of these fields provides an example of
Lagrangian field theory, the action functional being given by
\begin{equation}
S \left( \bar{\psi}, \psi; A, Z\right) \;:= \int d^{2n}x~\bar{\psi}(x) 
D_{A,Z} \psi(x)~, 
\label{1.25}
\end{equation}
where the covariant Dirac operator is 
\begin{equation}
D_{A,Z} \;=\; i \gamma^\mu \left( \partial_\mu - i A_\mu - i Z_\mu
  \gamma \right)~,
\label{1.26}
\end{equation}
with $\gamma ~(=$ ``$\gamma^5$'') as in eq.~(\ref{1.1}). The fields
$A_\mu$ and $Z_\mu$ are arbitrary external fields (i.e., they are not
quantized, for the time being). We define the effective action,
$S_{\rm eff} (A,Z)$, by
\begin{equation}
e^{i S_{\rm eff} (A,Z)} \; :=\; {\rm const} \int {\mathcal D} \psi
{\mathcal D} \bar{\psi}\; e^{i S(\bar{\psi},\psi;A,Z)} \ ,
\label{1.27}
\end{equation}
where the constant is chosen such that $S(A=0, Z=0)=0$, and $\hbar$
and $c$ have been set to 1. After Wick rotation,
\begin{equation}
t = x^0 \to - ix^0,~ A_0 \to iA_0, ~Z_0 \to i Z_0, ~\gamma^0 \to -
i\gamma^0~, 
\label{1.28}
\end{equation}
eq.~(\ref{1.27}) reads
\begin{equation}
e^{- S_{\rm eff}^E (A,Z)} \;=\; \left[ \int {\mathcal D} \psi
  {\mathcal D} \bar{\psi}\; e^{- S^E (\bar{\psi},\psi; A, Z)}
\right]_{\rm ren}
\label{1.29}
\end{equation}
where the integral on the R.S. is interpreted as 
a renormalized Gaussian {\it Berezin integral}. Thus
\begin{equation}
e^{-S_{\rm eff}^E(A,Z)}\;=\; {\rm det}_{\rm ren} \left(
  D_{A,Z}\right)~, 
\label{1.30}
\end{equation}
where, after Wick rotation,
\[
D_{A,Z} \;=\; i \,\gamma^\mu \left( \partial_\mu - i\,A_\mu - i\,Z_\mu 
  \gamma\right) 
\]
is an anti-hermitian elliptic operator, and the subscripts ``ren'' indicate
that (for $n \geq 2$) a multiplicative renormalization must be made.

The effective action $S_{\rm eff}^E (A,Z)$ is the generating function
for the Euclidian Green functions of the vector- and axial vector
currents. At non-coinciding arguments, 
\begin{eqnarray}
&&{} \langle {\mathcal J}^{\mu_1} (x_1) \ldots \widetilde{{\mathcal
    J}}^{\nu_1} (y_1) \ldots \rangle_{A,Z}^c \nonumber\\
&&{}~~~~= (- iq) \; \frac{\delta}{\delta A_{\mu_1}
  (x_1)}\;\ldots\;(-iq)\; \frac{\delta}{\delta Z_{\nu_1}(y_1)} \;
\ldots S_{\rm eff}^E (A,Z)~,
\label{1.31}
\end{eqnarray}
where $q$ is the electric charge, and $\langle (\cdot)\rangle_{A,Z}^c$ 
denotes a {\it connected} expectation value. 

We should like to understand how $S_{\rm eff}^E (A,Z)$ changes under
the gauge transformations
\begin{equation}
A_\mu \to A_\mu + \partial_\mu \chi~,~~Z_\mu \to Z_\mu + \partial_\mu
\alpha~. 
\label{1.32}
\end{equation}
Following Fujikawa [2], we perform a phase transformation and a chiral 
rotation of $\psi$ and $\bar{\psi}$ under the integral on the R.S. of
eq.~(\ref{1.29}). We set 
\begin{equation}
\psi' (x) \;=\; e^{i\left( \chi (x) +\alpha (x) \gamma\right)}\, \psi
(x)~,~~ \bar{\psi}' (x) \;=\; \bar{\psi}(x) \; e^{-i\left( \chi (x) -
    \alpha (x)\gamma\right)} ~. 
\label{1.33}
\end{equation}
Then
\begin{equation}
S^E\left( \bar{\psi}', \psi'; A+d\chi, Z+d \alpha\right) \;=\; S^E
\left( \bar{\psi}, \psi; A,Z\right)~,
\label{1.34}
\end{equation}
where $d\chi$ denotes the gradient, $(\partial_\mu \chi)$, of
$\chi$. Next, we must determine the Jacobian, $J$, of the
transformation (\ref{1.33}),
\begin{equation}
{\mathcal D} \,\bar{\psi}'\;{\mathcal D}\, \psi' \;=:\; J\, {\mathcal
  D}\, \bar{\psi} \,{\mathcal D}\, \psi~.
\label{1.35}
\end{equation}
Obviously, phase transformations,
\[
\psi' \;=\; e^{i\chi}\,\psi~,~~\bar{\psi}' \;=\; \bar{\psi}\;
e^{-i\chi} 
\]
have Jacobian $J=1$. However, this may {\it not} be so for chiral
rotations. Formally, under chiral rotations, the Jacobian turns out to 
be 
\begin{equation}
J \;=\; {\rm exp}\, \left[ 2 i\, {\rm Tr} \left( \alpha
    \gamma\right)\right]~. 
\label{1.36}
\end{equation}  
The problem with the R.S. of (\ref{1.36}) is that, a priori, it is
ill-defined. Let us assume that non-compact Euclidian space-time is
replaced by a $2n$-dimensional sphere. Then $D_{A,Z}$ has discrete
spectrum, with eigenvalues $i\,\lambda_m$ corresponding to eigenspinors
$\psi_m (x),~m \in {\mathbb Z}$. Formally,
\[
{\rm Tr}\, (\alpha \gamma) \;=\; \sum_m \int d^{2n}x\;\alpha
(x)\;\psi_m^* (x) \gamma \; \psi_m (x)~.
\]
We regularize the R.S. by replacing it by
\begin{equation}
\sum_m e^{-\left( \lambda_m^2 \big / M^2\right)} \int d^{2n}x\; \alpha 
(x)\; \psi_m^* (x) \gamma\; \psi_m (x)
\label{1.37}
\end{equation}
and, afterwards, letting $M\to\infty$. Expression (\ref{1.37}) is nothing 
but 
\begin{equation}
{\rm Tr}\,\left( \alpha\, \gamma\;e^{ \left( D_{A,Z}^2 \big/
      M^2\right)}\right)~.
\label{1.38}
\end{equation}

From Alvarez-Gaum\'e's calculations [3] concerning the index theorem,
for example, we infer that 
\begin{equation}
\lim_{M\to\infty} {\rm Tr}\,\left( \alpha\,\gamma\;e^{\left( D_{A,Z}^2
      \big/ M^2\right)} \right) \;=\;- \int d^{2n}x\;\alpha (x)
\;{\mathcal A} (x)~,
\label{1.39}
\end{equation}
where ${\mathcal A}(x)$ is the {\it index density} described more
explicitly below. From (\ref{1.39}) and (\ref{1.36}) we obtain that 
\begin{equation}
J \;=\; {\rm exp}\,\left[ - 2i \int d^{2n} x\;\alpha (x) \;{\mathcal
    A} (x) \right]~.
\label{1.40}
\end{equation}
With (\ref{1.34}), (\ref{1.35}) and (\ref{1.29}), eq.~(\ref{1.40})
yields
\begin{equation}
S_{\rm eff}^E \left( A + d \chi, Z + d \alpha\right) \;=\; S_{\rm
  eff}^E (A,Z) \,-\, 2 i \int d^{2n} x\; \alpha (x) \; {\mathcal A}
(x)~.
\label{1.41}
\end{equation}
When combined with (\ref{1.31}) eq.~(\ref{1.41}) is seen to yield
\begin{eqnarray}
&&{} \left[ \delta\,S_{\rm eff}^E \left( A+ d\chi,0\right) \big/
  \delta \chi (x)\right]_{\chi=0} \nonumber\\
&&{}~~~~~=\; \partial_\mu \left( \delta\,S_{\rm eff}^E (A,0) \big/
  \delta\,A_\mu (x)\right)\nonumber\\
&&{}~~~~~=\; \frac i q \ \partial_\mu \langle {\mathcal J}^\mu
(x)\rangle_A \;=\; 0~,
\label{1.42}
\end{eqnarray}
and
\begin{eqnarray}
&&{} \left[ \delta\,S_{\rm eff}^E \left( A, Z + d \alpha\right) \big/
  \delta \alpha (x) \right]_{Z=\alpha=0} \nonumber\\
&&{}~~~~~=\; \partial_\mu \left(\left[ \delta\,S_{\rm eff} \left(
      A,Z\right) \big/ \delta\,Z_\mu (x)\right]_{Z=0}\right)
\nonumber\\
&&{}~~~~~=\; \frac i q \ \partial_{\mu}\langle \widetilde{{\mathcal J}}^\mu
(x)\rangle_A \;=\; - 2 i \, {\mathcal A} (x)~,
\label{1.43}
\end{eqnarray}
i.e.,
\begin{equation}
\partial_\mu \langle {\mathcal J}^\mu (x)\rangle_A\;=\;0~,~~
\partial_\mu 
\langle \widetilde{{\mathcal J}}^\mu (x)\rangle_A \;=\; - 2
q\,{\mathcal A} (x)~.
\label{1.44}
\end{equation}

Introducing the chiral currents
\begin{equation}
{\mathcal J}_\ell^\mu \;:=\; {\mathcal J}^\mu -
  \widetilde{{\mathcal J}}^\mu~,~~{\mathcal J}_r^\mu \;:=\;
 {\mathcal J}^\mu + \widetilde{{\mathcal
      J}}^\mu~, 
\label{1.45}
\end{equation}
where ${\mathcal J}_{\ell/r}^\mu$ is the current of
left-handed/right-handed fermions, we see that (\ref{1.44}) is
equivalent to 
\begin{equation}
\partial_\mu \langle {\mathcal J}_\ell^\mu (x)\rangle_A \;=\;
2\,q\,{\mathcal A}(x)~,~~ \partial_\mu \langle {\mathcal J}_r^\mu
(x)\rangle_A \;=\; - 2\,q\,{\mathcal A} (x)~.
\label{1.46}
\end{equation}

Locally, we can solve the equation
\begin{equation}
\delta\, \omega \left( x;A\right) \;=\; {\mathcal A} (x)~,
\label{1.47}
\end{equation}
where $\delta$, the co-differential, is the dual of exterior
differentiation $d$, the solution 
$\omega(\,\cdot\,;\,A)$ being a 1-form. The 1-form
$\omega(\,\cdot\,;\,A)$ is, however, not gauge-invariant. We may now
define modified currents,
\begin{equation}
\widehat{{\mathcal J}}_{\ell/r}^\mu (x) \;=\; {\mathcal
  J}_{\ell/r}^\mu (x) \;\mp\; 2\,q\,\omega^\mu (x;A)~.
\label{1.48}
\end{equation}
They are not gauge-invariant, but, according to eqs.~(\ref{1.47}),
(\ref{1.48}), they are conserved, i.e.,
\begin{equation}
\partial_\mu \ \widehat{{\mathcal J}}_{\ell/r}^\mu (x) \;=\; 0~.
\label{1.49}
\end{equation}
Passing to the operator formulation of quantum field theory (i.e.,
undoing the Wick rotation (\ref{1.28}), which amounts to
Osterwalder-Schrader reconstruction), the conserved currents
$\widehat{{\mathcal J}}_{\ell/r}^\mu$ give rise to {\it conserved
  charges}, 
\begin{equation}
N_{\ell/r} \;:= \int d\underline{x}\; \widehat{{\mathcal
    J}}_{\ell/r}^0 (t, \underline{x})
\label{1.50}
\end{equation}
which (for gauge-transformations continuous at infinity) {\it are}
gauge-invariant.

We should like to determine the {\it equal-time commutators} of the
({\it gauge-invariant}) currents 
${\mathcal J}_{\ell/r}^\mu (x)$. Let ${\mathcal V}$ denote the affine
space of configurations of external electromagnetic vector potentials, 
$A$, corresponding to {\it static} electromagnetic fields. We consider 
the Hilbert bundle, ${\mathcal H}$, over ${\mathcal V}$ whose fibre,
${\mathcal F}_A$, at a point $A \in {\mathcal V}$ is the Fock space of 
state vectors of free, chiral (e.g., left-handed) fermions coupled to
the vector potential $A$. Then ${\mathcal H}$ carries a projective
representation, $U$, of the group ${\mathcal G}$ of {\it
  time-independent} electromagnetic gauge transformations, 
\[
g\;=\;\left( g^\chi (x)\right)~,~~g^\chi (x)\;=\;e^{i\chi (x)}~,~~
\chi (x)\;=\; \chi (\underline{x})~~({\rm indep.~of~} t)~,
\]
with the following properties: 

\begin{description}
\item[(i)]~~~~~$U (g)~:~ {\mathcal F}_A \;\longrightarrow\; {\mathcal
  F}_{A+d\chi}$ \ ,\\
\end{description}
and, on the fibre ${\mathcal F}_{A+d\chi}$~,
\begin{description}
\item[(ii)]~~~~~$U(g)\,\psi (x;A)\bigm|_{{\mathcal F}_A}\;
  U(g)^{-1}\;=\; e^{i\chi (x)}\; \psi (x; A+d\chi)$ ,
\end{description}
where $\psi (x;A)$ is the Dirac spinor field acting on ${\mathcal
  F}_A$; (and similarly for $\bar{\psi} (x;A)$). The generator,
$G(\chi)$, of the gauge transformation $U(g^\chi (\cdot))$ is given by
\[
G (\chi) \;= \int d\underline{x} \;\chi (\underline{x})\; G(x)~,
\]
where 
\begin{equation}
G(x) \;=\; - i\, \underline{\nabla}~\cdot~ \frac{\delta}{\delta
  \,\underline{A}\,(x)}~+~\frac 1 q \ {\mathcal J}_\ell^0 (x;A)~. 
\label{1.51}
\end{equation}
Here
\[
\underline{\nabla}~\cdot~\frac{\delta}{\delta\;\underline{A}} \ = \
\sum_{j=1}^{2n-1} \partial_j~\cdot~\frac{\delta}{\delta\, A_j} \ .
\]
{\it Locally}, the (phase) factor of the projective representation $U$ 
of ${\mathcal G}$ can be made {\it trivial} by redefining the
generators $G$:
\begin{equation}
G(x)\;\longrightarrow\; \widehat{G}(x) \;:=\; - i\,
\underline{\nabla}~\cdot~ \frac{\delta}{\delta \,\underline{A} (x)}~+~ 
\frac 1 q \ \widehat{{\mathcal J}}_\ell^0 (x;A)~.
\label{1.52}
\end{equation}
The operators $\widehat{G}(x)$ generate a {\it representation} of the
group ${\mathcal G}$ of gauge transformations on ${\mathcal H}$ iff
\begin{equation}
\left[ \widehat{G}\left( t,\underline{x}\right)~,~~ \widehat{G}\left(
    t,\underline{y}\right)\right] \;=\; 0
\label{1.53}
\end{equation}
for all times $t$. That (\ref{1.52}) is the right choice of generators 
compatible with (\ref{1.53}) follows, heuristically, from the fact
that $\widehat{{\mathcal J}}_\ell^\mu (x;A)$ is a {\it conserved}
current.

Because the current ${\mathcal J}_\ell^\mu (x;A)$ is gauge-invariant,
we have that
\begin{equation}
\left[ \underline{\nabla}~\cdot~\frac{\delta}{\delta\,\underline{A}
    (x)}~,~~ {\mathcal J}_\ell^0 \left( y;A\right)\right] \;=\; 0~.
\label{1.54}
\end{equation}
Thus, using (\ref{1.48}), (\ref{1.54}) and (\ref{1.53}), we find that 
\begin{eqnarray}
0 &\displaystyle\mathop{=}^!& \left[ \widehat{G} \left(
    t,\underline{x}\right)~,~~\widehat{G} \left( t, \underline{y}\right)\right]
\nonumber\\
&=& \frac{1}{q^2} \ \left[ {\mathcal J}_\ell^0 \left( t,
    \underline{x}\right)~,~~{\mathcal J}_\ell^0 \left( t,
  \underline{y}\right)\right]~-~ 2\,i\,\underline{\nabla}~\cdot~
\frac{\delta}{\delta\,A (t, \underline{x})} \ \omega^0 \left( t,
  \underline{y}; A\right) \nonumber\\
&&{}~~~~~~~~~~~
+\;2\,i\,\underline{\nabla}~\cdot~\frac{\delta}{\delta\,A
  (t,\underline{y})} \ \omega^0 \left( t, \underline{x}; A\right)~. 
\label{1.55}
\end{eqnarray}
This equation determines the anomalous commutator
\begin{equation}
\left[ {\mathcal J}_\ell^0 \left( t, \underline{x}\right)~,~~{\mathcal 
    J}_\ell^0 \left( t, \underline{y}\right)\right] \;=\; \left[
  \widehat{{\mathcal J}}_\ell^0 \left( t, \underline{x}\right)~,~~
\widehat{{\mathcal J}}_\ell^0 \left( t, \underline{y}\right)\right]~. 
\label{1.56}
\end{equation}
Of course, our arguments are heuristic, but, hopefully,
provide a reasonably clear idea about the origin of anomalous
commutators. A more erudite, mathematically clean derivation of
(\ref{1.55}) can be based on an analysis of the {\it cohomology} of
${\mathcal G}$; see e.g. [4].

In order to arrive at explicit versions of eqs.~(\ref{1.46}),
(\ref{1.47}) and (\ref{1.55}) in various even dimensions, we must know 
the explicit expressions for the index density ${\mathcal A} (x)$ and
the one-form $\omega (x; A)$. We shall not have any occasion to
consider systems coupled to a non-trivial chiral gauge field $Z$. We
therefore set $Z=0$. Then, in two space-time dimensions, 
\begin{equation}
{\mathcal A}(x) \;=\; - \ \frac{1}{4\pi} \ E(x)~,
\label{1.57}
\end{equation}
by comparison of (\ref{1.45}) with (\ref{1.20}), and, by (\ref{1.48})
and (\ref{1.57}), 
\begin{equation}
\omega^\mu (x;A) \;=\; - \ \frac{1}{4\pi} \  \varepsilon^{\mu\nu}\,
A_\nu (x)~,
\label{1.58}
\end{equation}
see also (\ref{1.23}) and (\ref{1.17}). In four space-time dimensions
\begin{eqnarray}
{\mathcal A} (x) &=& - \ \frac{1}{32\,\pi^2}~*~ \left( F \wedge
  F\right) (x) \nonumber\\
&=& - \ \frac{1}{32\,\pi^2}~ F_{\mu\nu} (x)~\widetilde{F}^{\mu\nu}
(x)~, 
\label{1.59}
\end{eqnarray}
where $\wedge$ denotes the exterior product and $*$ the ``Hodge
dual''. By eq.~(\ref{1.47}), 
\begin{equation}
\omega^\mu \left( x;A\right) \;=\; -~\frac{1}{32\,\pi^2} \
\varepsilon^{\mu\nu\lambda\rho}~A_\nu (x)~F_{\lambda\rho} (x)~.
\label{1.60}
\end{equation}
Thus eqs.~(\ref{1.47}) read
\begin{equation}
\partial_\mu \langle {\mathcal J}_{\ell/r}^\mu (x)\rangle_A \;=\; \mp
\ \frac{q}{32\,\pi^2}~*~\left( F \wedge F\right) (x)~,
\label{1.61}
\end{equation}
and, from eqs.~(\ref{1.55}), (\ref{1.56}) and (\ref{1.60}), we
conclude that
\begin{eqnarray}
&&{} \left[ {\mathcal J}_{\ell/r}^0 \left( t,
    \underline{x}\right)~,~~{\mathcal J}_{\ell/r}^0 \left(
    t,\underline{y}\right)\right] \;=\; \left[ \widehat{{\mathcal
      J}}_{\ell/r}^0 \left( t,
    \underline{x}\right)~,~~\widehat{{\mathcal J}}_{\ell/r}^0 \left(
    t, \underline{y}\right)\right] \nonumber \\
&&{}~~~~~=\; \pm~i~\frac{q^2}{4\pi^2} \ \left( \underline{B}
  (\underline{x},t)\,\cdot\,\underline{\nabla}\right)~\delta~\left(
  \underline{x} - \underline{y}\right)~,
\label{1.62}
\end{eqnarray}
a well known result; see [1].

The key fact reviewed in this section, from which all other results
can be derived, is eq.~(\ref{1.41}), i.e.,
\begin{equation}
S_{\rm eff}^E \left( A + d\chi,~Z+d\alpha\right) \;=\; S_{\rm eff}^E
\left( A, Z\right) - 2\,i \int d^{2n}x~\alpha (x)~{\mathcal A} (x)~. 
\label{1.63}
\end{equation}

In order to describe a system in which only the left-handed fermions
are charged, while the right-handed fermions are {\it neutral}, one
may just set
\begin{equation}
A \;=\; -\, Z \;=\;  a
\label{1.64}
\end{equation}
in eq.~(\ref{1.63}), where $a$ is the electromagnetic vector potential 
to which the left-handed fermions are coupled; see (\ref{1.25}),
(\ref{1.26}) and (\ref{1.46}). Denoting the effective action of this
system by $W_\ell (a)$, we find from (\ref{1.63}) and (\ref{1.64})
that 
\begin{equation}
W_\ell \left( a+d\chi\right) \;=\; W_\ell (a) +\, 2\,i \int d^{2n}
x\,\chi (x) \,{\mathcal A} (x)~.
\label{1.65}
\end{equation}
Similarly,
\begin{equation}
W_r \left( a+d\chi\right) \;=\; W_r (a) - 2\,i \int d^{2n}x\,\chi (x)
\,{\mathcal A} (x)~,
\label{1.66}
\end{equation}
for charged {\it right-handed} fermions.

Eqs.~(\ref{1.65}) and (\ref{1.66}) show that a theory of massless {\it
  chiral} fermions coupled to an external electromagnetic field is
{\it anomalous}, in the sense that it fails to be gauge-invariant. But 
let us imagine that space-time, ${\mathbb R}^{2n}$, is the boundary of 
a $(2n+1)$-dimensional half-space $M$, (i.e., $\partial M$ = physical
space-time $\cong {\mathbb R}^{2n}$). Let $A$ denote a smooth
U(1)-gauge potential on $M$ which is continuous on $\partial M$, with
\begin{equation}
A\bigm|_{\partial M} \;=\; a~.
\label{1.67}
\end{equation}
Let $\omega^{2n+1} (\cdot; A)$ denote the usual Chern-Simons
$(2n+1)$-form on $M$. The Chern-Simons action on $M$ is defined by 
\begin{equation}
S_{CS}(A) \;:=\; i \int_M \omega^{2n+1} \left( \xi; A\right)~,
\label{1.68}
\end{equation}
where $\xi$ denotes a point in $M$. It should be recalled that
\begin{equation}
\omega^{2n+1} \left( \cdot ; A+d\chi\right) \;=\; \omega^{2n+1} (\cdot 
; A) \,+\, d\chi \wedge \left( * \,{\mathcal A}\right)~.
\label{1.69}
\end{equation}
Since $d(*\,{\mathcal A})=0~$, \ $d\chi \wedge (*\,{\mathcal A}) = d
\left(\chi\left( *\,{\mathcal A}\right)\right)~$, and hence, by
Stokes' theorem,
\begin{eqnarray}
S_{CS} \left( A+ d\chi\right)
&=& S_{CS} (A) + i \int_{\partial M} \chi (x)\, (*\,{\mathcal
  A}) (x) \nonumber\\
&=& S_{CS} (A) + i \int_{\partial M} d^{2n} x\, \chi (x) \,
{\mathcal A} (x)~.
\label{1.70}
\end{eqnarray}
It follows that
\begin{equation}
W_{\ell/r} (a)\;\mp\; S_{CS} (A)~~{\rm is~} gauge-invariant.
\label{1.71}
\end{equation}

This result has a $(2n+1)$-dimensional interpretation (see [5]): 
Consider a
(parity-violating) theory of massive, charged fermions described by
$2^n$-component Dirac spinors on a $(2n+1)$-dimensional space-time $M$ 
with non-empty boundary $\partial M$. These fermions are minimally
coupled to an external electromagnetic vector potential $A$. We impose 
some anti-selfadjoint spectral boundary conditions on the
$(2n+1)$-dimensional, covariant Dirac operator $D_A$. The action of
the system is given by
\begin{equation}
S \left( \bar{\psi},\psi;A\right) \;:=\; \int_M d^{2n+1}
\xi\,\bar{\psi} (\xi) \left( D_A+m\right)\, \psi (\xi)~,
\label{1.72}
\end{equation}
where $m$ is the bare mass of the fermions. The effective action of
the system is defined by
\begin{eqnarray}
e^{- S_{\rm eff}^E(A)} &:=& \left[ \int {\mathcal D} \psi \,{\mathcal D}
  \bar{\psi}\; e^{-S^E \left(\bar{\psi},\psi;A\right)}\right]_{\rm ren}
\nonumber\\
&=& {\rm det}_{\rm ren} \left( D_A + m\right)~,
\label{1.73}
\end{eqnarray}
where the subscript ``ren'' indicates that renormalization may be
necessary to define the R.S. of (\ref{1.73}). Actually, for $n=1$, no
renormalization is necessary; but, for $n=2$, e.g. an infinite charge
renormalization must be made. It turns out that, for $n=1$ and $n=2$
(after renormalization),
\begin{equation}
S_{\rm eff}^E (A) \;=\; W_{\ell/r} \left( A\bigm|_{\partial
    M}\right)\,\mp\, S_{CS} (A)\,+\, {\mathcal O}\, \left( \frac 1 m
\right) ~,
\label{1.74}
\end{equation}
up to a Maxwell term depending on renormalization conditions, where
the correction terms are manifestly gauge-invariant; see
[5,6]. (Whether the R.S. of (\ref{1.74}) involves $W_\ell$ or $W_r$
depends on the definition of $D_A$).

The physical reason underlying the result claimed in eq.~(\ref{1.74})
is that, in a system of massive fermions described by $2^n$-component
Dirac spinors confined to a space-time $M$ with a non-empty,
$2n$-dimensional boundary $\partial M$, there are {\it massless,
  chiral fermionic surface modes} propagating along $\partial M$.

This completes our heuristic review of aspects of the chiral abelian
anomaly that are relevant for the physical applications to be
discussed in subsequent sections. The abelian anomaly is, of course,
but a special case of the general theory of anomalies involving also
non-abelian, gravitational, global, \dots anomalies. In recent years,
this theory has turned out to be important in connection with the
theory of branes in string theory and 
with understanding aspects of $M$-theory. But, in this review, such
applications will not be described.

In Sect.~2, we describe physical systems, important features of which
can be understood as consequences of the two-dimensional chiral
anomaly: incompressible (quantum) Hall fluids and ballistic wires.

In Sect.~3, we describe degrees of freedom in four dimensions which
may play an important r\^ole in the generation of seeds for cosmic
magnetic fields in the very early universe.
This will turn out to be connected with the four-dimensional chiral
anomaly. 

In Sect.~4, a brief review of the theory of ``transport in thermal
equilibrium through gapless modes'' developed in [7] is presented.

In Sects.~5 and 6, we combine the results of this section with those
in Sect.~4 to derive physical implications of the chiral anomaly for
the systems introduced in Sects.~2 and 3.

Some conclusions and open problems are described in Sect.~7.

\setcounter{equation}{0}
\section{Quantized conductances}

The original motivation for the work described in this review has been 
to provide simple and conceptually clear explanations of various
formulae for quantized conductances, which have been encountered in
the analysis of experimental data. Here are some typical examples.

\vspace{.3cm}

\noindent
{\bf Example 1.} Consider a quantum Hall device with, e.g., an annular 
(Corbino) geometry. Let $V$ denote the voltage drop in the radial
direction, between the inner and the outer edge, and let $I_H$ denote
the total Hall current in the azimuthal direction. The Hall
conductance, $G_H$, is defined by
\begin{equation}
G_H \;=\; I_H/V~.
\label{2.1}
\end{equation}
One finds that {\it if the longitudinal resistance vanishes} (i.e., if 
the two-dimensional electron gas in the device is ``incompressible'')
then $G_H$ is a {\it rational multiple} of $\frac{e^2}{h}$, i.e., 
\begin{equation}
G_H \;=\; \frac n d \ \cdot \ \frac{e^2}{h}~,~~n\;=\;0,1,2,\ldots,
~~d\;=\; 1,2,3,\ldots~.
\label{2.2}
\end{equation} 
In (\ref{2.2}), $e$ denotes the elementary electric charge and $h$
denotes Planck's constant. Well established Hall fractions, $f_H :=
\frac n d$, in the range $0 < f_H \leq 1$ are listed in Fig.~1; (see [8]; 
and [9, 10, 11] for general background). 

\vspace{.3cm}

\noindent
{\bf Example 2.} In a ballistic (quantum) wire, i.e., in a pure, very
thin wire without back scattering centers, one finds that the
conductance $G_W = I/V$ ($I$: current through the wire, $V$: voltage
drop between the two ends of the wire) is given by
\begin{equation}
G_W\;=\;2N~\frac{e^2}{h}~,~~N\;=\;0,1,2,\ldots,
\label{2.3}
\end{equation}
under suitable experimental conditions (``small'' $V$, temperature not 
``very small'', ``adiabatic gates''); see [12, 13].

\vspace{.3cm}

\noindent
{\bf Example 3.} In measurements of {\it heat conduction} in quantum
wires, one finds that the heat current is an {\it integer multiple} of 
a ``fundamental'' current which depends on the temperatures of the two 
heat reservoirs at the ends of the wire.

\begin{figure}[t]
\begin{center}
\psfig{file=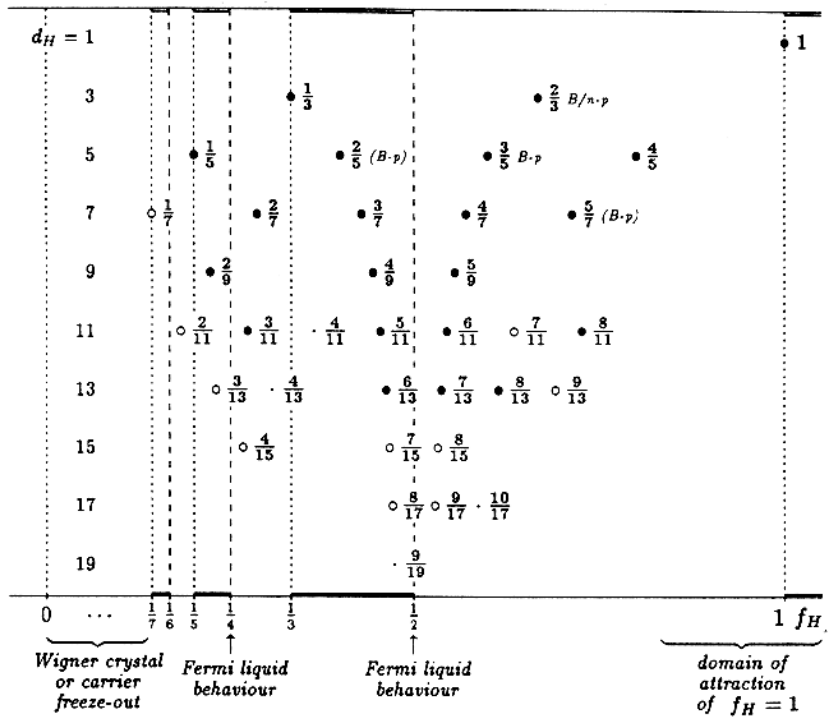} 
\caption{  \label{fig:1}
  Observed Hall fractions $\sigma_H=n_H/d_H$\ 
  in the interval $0<\sigma_H\leq1$. }
\end{center}
\end{figure}

If electromagnetic waves are sent through an ``adiabatic hole''
connecting two half-spaces one approximately finds an ``integer
quantization'' of electromagnetic energy flux.

Our task is to attempt to provide a theoretical explanation of these
remarkable experimental discoveries; hopefully one that enables us to
predict further related effects.

Conductance quantization is observed in a rather wide temperature
range. It appears that it is only found in systems without dissipative 
processes. When it is observed it is insensitive to small changes in
the parameters specifying the system and to details of sample
preparation; i.e., it has {\it universality properties}. --- It will
turn out that the key feature of systems exhibiting conductance
quantization is that they have {\it conserved chiral charges}; (such
conservation laws will only hold approximately, i.e., in slightly
idealized systems). Once one has understood this point, the right
formulae follow almost automatically, and one arrives at natural
generalizations.

In order to give a first indication how the effects described here
might be related to the two-dimensional chiral anomaly, we consider
Example~1, the quantum Hall effect, in more detail. 
For readers not familiar with this remarkable effect [14], we summarize
some of its key features.

A quantum Hall fluid (QHF) is an interacting electron gas confined to
some domain in a two-dimensional plane (an interface between a
semiconductor and an insulator, with compensating background charge)
subject to a constant magnetic field $\vec{B}^{(0)}$ transversal to
the confinement plane. Among experimental control parameters is the
filling factor, $\nu$, defined by
\[
\nu \;=\; \frac{n^{(0)}}{B^{(0)} \big/ \left( \frac{h\,c}{e}\right)}
\]
where $n^{(0)}$ is the (constant) electron density, $B^{(0)}$ is the
component of the magnetic field $\vec{B}^{(0)}$ perpendicular to the
plane of the fluid, and $\frac{h\,c}{e}$ is the quantum of magnetic
flux. The filling factor $\nu$ is dimensionless.

Transport properties of a QHF in an external electric field (of small
frequency) are described by the equation
\begin{equation}
\underline{J} \left(t, \underline{x}\right) \;=\; 
{~\sigma_L~~~\sigma_H \choose -\sigma_H~~\sigma_L} ~~\underline{E}
\left( t, \underline{x}\right)~,
\label{2.4}
\end{equation}
where $\underline{x}$ is a point in the sample, $\underline{J}$ is the 
bulk electric current parallel to the sample plane and $\underline{E}$ 
is the component of the external electric field parallel to the sample 
plane. Furthermore, $\sigma_L$ denotes the longitudinal conductivity,
and $\sigma_H$ is the transverse -- or Hall conductivity. In two
dimensions, conductances and conductivities have the same dimension of 
[(charge)$^2$/action], and it is not difficult to see that
\begin{equation}
G_H \;=\; \sigma_H~.
\label{2.5}
\end{equation}

Experimentally, one observes that the longitudinal conductivity,
$\sigma_L$, vanishes when the filling factor $\nu$ belongs to certain
small intervals [9], a sign that there are no dissipative processes
in the fluid. Such a QHF is called ``{\it incompressible}'', for
reasons explained below. Furthermore, on every interval of $\nu$ where 
$\sigma_L$ vanishes, the Hall conductivity $\sigma_H$ is a rational
multiple of $\frac{e^2}{h}$~, as claimed in (\ref{2.2}). 

Next, we recall the basic equations of the electrodynamics of an
incompressible QHF; see [8]. It is useful to combine the
two-dimensional space of the fluid and time to a three-dimensional
space-time. The electromagnetic field tensor of the system is given by
\begin{equation}
F_{\mu\nu} \;=\; \ \left(
\begin{array}{lll}
~~0 &E_1 &E_2\\
-E_1~~ &~0 & B\\
-E_2 &-B~~ &0
\end{array}
\right)~,
\label{2.6}
\end{equation}
where $E_1$ and $E_2$ are the components of an external electric field 
in the plane of the sample, and $B$ is the component of an external
magnetic field, $\vec{B}$, perturbing the constant field
$\vec{B}^{(0)}$ perpendicular to the sample plane; $\left(
  \vec{B}_{\rm total} = \vec{B}^{(0)} + \vec{B}\right)$.

We define $J^0(x)$ to denote the sum of the electron charge density in 
the space-time point $x=(t,\underline{x})$ and the uniform background
charge density $en^{(0)}$. We set $J^\mu = (J^0,\underline{J})$. 

From the three-dimensional homogeneous Maxwell equations (Faraday's
law),
\begin{equation}
\partial_\mu\,F_{\nu\lambda}\;+\;\partial_\nu\,F_{\lambda\mu}\;+\;
\partial_\lambda\,F_{\mu\nu} \;=\; 0~,
\label{2.7}
\end{equation}
 the continuity equation for the electric current density
 (conservation of electric charge),
\begin{equation}
\partial_\mu\,J^\mu \;=\; 0~,
\label{2.8}
\end{equation}
and from the transport equation (\ref{2.4}) with $\sigma_L=0$, it
follows [8] that
\begin{equation}
J^0 \;=\; \sigma_H\, B~.
\label{2.9}
\end{equation}
Equations (\ref{2.4}), for $\sigma_L=0$, and (\ref{2.9}) can be
combined to the equation
\begin{equation}
J^\mu \;=\; \sigma_H\;\varepsilon^{\mu\nu\lambda}\;F_{\nu\lambda}
\label{2.10}
\end{equation}
of Chern-Simons electrodynamics, [5].
Eqs.~(\ref{2.10}) describe the response of an incompressible QHF to an 
external electromagnetic field (perturbing the constant magnetic field 
$\vec{B}^{(0)}$).

Unfortunately, eqs.~(\ref{2.10}) are compatible with the continuity
equation (\ref{2.8}) for $J^\mu$ only if $\sigma_H$ is constant
throughout space-time. But realistic samples have a finite extension.

The finite extension of the sample, confined to a space-time region
$\Omega = D \times {\mathbb R}$, where $D$ is e.g. a disk or an
annulus, is taken into account by setting the Hall conductivity
$\sigma_H (\cdot)$ to zero outside $\Omega$, i.e.,
\begin{equation}
\sigma_H (\xi) \;=\; \sigma_H\;\chi_\Omega (\xi)~,
\label{2.11}
\end{equation}
for $\xi \in {\mathbb R}^3$, where $\sigma_H$ is the (constant) value
of the Hall conductivity inside the sample, and $\chi_\Omega$ is the
characteristic function of $\Omega$. Taking the divergence of
eq.~(\ref{2.10}), we get that
\begin{equation}
\partial_\mu\;J^\mu \;=\; \sigma_H\;
\varepsilon^{\mu\nu\lambda}\;\left(
  \partial_\mu\,\chi_\Omega\right)~F_{\nu\lambda}~, 
\label{2.12}
\end{equation}
i.e., $\partial_\mu J^\mu$ {\it fails to vanish} on the boundary,
$\partial D$, of the sample. However, conservation of electric charge
is a fundamental law of nature 
for closed systems. Thus, there must be an electric current,
$J_{\partial \Omega}$, localized on the boundary $\partial \Omega$ of
the sample space-time such that the {\it total} electric current
\begin{equation}
J_{\rm total}^\mu \;=\; J^\mu + J_{\partial\Omega}^\mu
\label{2.13}
\end{equation}
satisfies the continuity equation. The boundary current
$J_{\partial\Omega}^\mu$ must be tangential to the boundary $\partial
\Omega$ of the sample space-time. Hence it determines a current
density, $I^\alpha$, on the (1+1)-dimensional space-time
$\partial\Omega$, where the index $\alpha$ refers to a choice of
coordinates on $\partial\Omega$. Eq.~(\ref{2.12}) and the continuity
equation for $J_{\rm total}^\mu$ then imply that
\begin{equation}
\partial_\alpha\,I^\alpha \;=\; -\, \sigma_H\;
\varepsilon^{\alpha\beta}\;F_{\alpha\beta}~. 
\label{2.14}
\end{equation}
This equation identifies $I^\alpha$ as an {\it anomalous} current.
Thus, there must be chiral modes (left-movers or right-movers,
depending on the orientation of $\Omega$ and the direction of the
external magnetic field) propagating along the boundary. They carry
the well known diamagnetic edge currents. If ${\mathcal J}_\ell^\alpha
$ (or ${\mathcal J}_r^\alpha$) denotes the corresponding
quantum-mechanical current operator then the edge current $I^\alpha$
is given by the quantum-mechanical expectation value,
$\langle {\mathcal J}_{\ell/r}^\alpha\rangle_A$, of ${\mathcal
  J}_\ell^\alpha$ (or ${\mathcal J}_r$). The currents ${\mathcal
  J}_\ell^\alpha$ have the anomalous commutators
\begin{equation}
\left[ {\mathcal J}_\ell^0 \left( t, \underline{x}\right), ~{\mathcal
    J}_\ell^0 \left( t, \underline{y}\right)\right] \;=\; \frac{i\,
\sigma_H}{2\pi}\; \delta' \left( \underline{x} - \underline{y}\right)~, 
\label{2.15}
\end{equation}
see eqs.~(\ref{1.21}) and (\ref{1.16}), and hence generate a chiral
$\hat{u}(1)$-current algebra with central charge given by
$\sigma_H$. 

We now return to the physics of the bulk of an incompressible QHF. The 
absence of dissipation $(\sigma_L = 0)$ in the transport of electric
charge through the bulk can be explained by the existence of a
mobility gap in the energy spectrum between the ground state energy of 
the QHF and the energies of extended, excited bulk states. This
property motivates the term ``incompressible'': It is not possible to
add an additional electron to, or subtract one from the fluid by
injecting only an arbitrarily small amount of energy. An important
consequence of incompressibility is that the total electric charge is
a good quantum number to label different sectors of physical states of 
an incompressible QHF (at zero temperature). 

We propose to study the bulk physics of incompressible QHF's in the
scaling limit, in order to describe the universal transport laws of such
fluids. For this purpose, we consider a QHF confined to a sample of
diameter $\propto \theta$, where $\theta$ is a dimensionless scale
factor. The {\it scaling limit} is the limit where $\theta\to\infty$,
with distances and time rescaled by a factor $\theta^{-1}$. In {\it
  rescaled coordinates}, the fluid is thus confined to a sample of
constant finite diameter.

The presence of a positive mobility gap in the system implies that, in 
the scaling limit, the effective theory describing an incompressible
QHF 
must be a ``{\it topological field theory}''. The states of a
topological field theory are indexed by static, pointlike sources
localized in the bulk and labelled by certain charge quantum numbers
which generate a fusion ring; see [8, 11].

It is not difficult [10] to find the effective action, $S_{\rm eff}
(A)$, in the scaling limit, where $A$ is the electromagnetic vector
potential  of the external
electromagnetic field $F_{\mu\nu}$, see eq.~(\ref{2.6}). A possible
starting 
point is eq.~(\ref{2.10}), relating the expectation value of the
electric current to the external electromagnetic field:
\begin{equation}
J^\mu (\xi) \;=\; \delta\,S_{\rm eff} (A) \big/ \delta\,A_\mu (\xi)
\;=\; \sigma_H\; \varepsilon^{\mu\nu\lambda}\; F_{\nu\lambda} (\xi)~.
\label{2.16}
\end{equation}
The solution of eq.~(\ref{2.16}) is
\begin{equation}
S_{\rm eff} (A) \;=\; \sigma_H\, S_{CS} (A) \;=\;
\frac{\sigma_H}{2} \int_\Omega d^3\xi\;
\varepsilon^{\mu\nu\lambda}\, A_\mu (\xi)\,\partial_\nu\,A_\lambda
(\xi)~,  
\label{2.17}
\end{equation}
i.e., $S_{\rm eff}$ is proportional to the Chern-Simons action
$S_{CS}$. The Chern-Simons action is not invariant under gauge
transformations of $A$ that do not vanish on the boundary $\partial
\Omega$ of the sample. Since electromagnetic gauge invariance is a
fundamental property of quantum-mechanical systems, eq.~(\ref{2.17})
for $S_{\rm eff}(A)$ must be corrected by a boundary term. Let $a$
denote the restriction of $A$ to the boundary $\partial \Omega$ of the 
sample. Then, as pointed out in eq.~(\ref{1.71}), the expression
$W_{\ell/r} (a) \mp S_{CS} (A)$ {\it is} gauge-invariant, where
$W_{\ell/r}(a)$ is the effective action of charged {\it chiral} modes
propagating along $\partial \Omega$. Thus, in the scaling limit,
\begin{equation}
S_{\rm eff} (A) \;=\; \sigma_H \left[ \mp W_{\ell/r} (a) +
  S_{CS} (A)\right]~,
\label{2.18}
\end{equation}
(depending on the sign of $\sigma_H$). It is well known that the
action $W_{\ell/r} (a)$ is the generating function for the connected
Green functions of the chiral current operators, ${\mathcal
  J}_{\ell/r}^\alpha$, on $\partial\Omega$, which generate a
$\hat{u}(1)$-current algebra. Formula (\ref{2.18}) plays an important
r\^ole in understanding the physics of incompressible quantum Hall
fluids. 

In the next section, we consider systems of massless chiral modes in
four-dimensional space-time, with physical properties some of which
are related to the four-dimensional chiral anomaly, and which may play 
a significant r\^ole in the {\it physics of the early universe.}

\setcounter{equation}{0}
\section{Branes, axions and charged fermions}

The very early universe is filled with a hot plasma of charged
leptons, quarks, gluons, photons, \dots~. At a time after the big bang
when the temperature $T$ is of the order of 80 $TeV$ chirality flips of
light charged leptons, in particular of right-handed electrons,
constitute a dynamical process slower than the expansion rate of the
universe. Thus, for $T\gtrsim$~80 $TeV$, the {\it chiral charges},
$N_\ell$ and $N_r$, defined in eq.~(\ref{1.50}) of Sect.~1, are
approximately conserved for electrons. They are related to an
approximate chiral symmetry of the electronic sector of the standard
model. Among other results, we shall attempt to show that if, in the
very early universe, the chemical potentials of left-handed and
right-handed electrons are different from each other, this may give
rise to the generation of large, cosmic magnetic fields, [15]; (see
also [7] for a similar, independent suggestion). This effect is, in a
sense explained in Sects.~4 and 6, an effect in {\it equilibrium
  statistical mechanics}. However, this is precisely what may make it
appear quite unnatural and implausible: The chiral charges, $N_\ell$
and $N_r$, are not really conserved; leptons are massive. The very
early universe is not really in an equilibrium state, and the chemical
potentials of left-handed and right-handed electrons neither have an
unambiguous meaning, {\it nor} would they be {\it space-} and {\it
  time-independent}. It may then be wrong, or, at least, misleading,
to invoke results from {\it equilibrium} statistical mechanics to
explore effects in the physics of the very early universe.

A way out from these difficulties can be found by seeking inspiration
from an analogy with the quantum Hall effect: Consider a quantum Hall
fluid (QHF), confined to a strip of macroscopic width $\ell$ in the
plane. If the QHF is {\it incompressible} then there are no light
(gapless) modes propagating through the bulk of the sample; but, as
shown in the last section, there are gapless, chiral modes propagating 
along the boundaries of the sample. Let $\Omega$ denote the 
space-time of the fluid; it is a slab of width $\ell$ in
three-dimensional Minkowski space. The two components of the boundary, 
$\partial \Omega$, of $\Omega$ are denoted by $\partial_+\Omega$,
$\partial_-\Omega$, respectively. As shown in the last section,
eq.~(\ref{2.18}), (see also [10] for more details) the effective
action of such an incompressible QHF (in the scaling limit) is given
by
\begin{equation}
S_{\rm eff} (A) \;=\; \sigma_H \left[ W_\ell \left(a_+\right) + 
  W_r \left( a_-\right) - S_{CS} (A)\right]~,
\label{3.1}
\end{equation}
(if the direction of the external magnetic field $\vec{B}^{(0)}$ is
chosen appropriately, given an orientation of $\Omega$). In
(\ref{3.1}), $A$ is an external electromagnetic vector potential on
$\Omega$, and
\begin{equation}
a_\pm  \;:=\; A\bigm|_{\partial_\pm \Omega}~,
\label{3.2}
\end{equation}
is the restriction of the 1-form $A$ to a component, $\partial_\pm
\Omega$, of the boundary of $\Omega$; $W_{\ell/r} (\cdot)$ is the
two-dimensional, anomalous effective action for charged, chiral
(left-moving, or right-moving, respectively) surface modes propagating 
along $\partial_+\Omega,~\partial_-\Omega$, respectively; and $S_{CS}
(\cdot)$ is the three-dimensional topological Chern-Simons action, see 
(\ref{2.17}). Many universal features of the quantum Hall effect can
be derived directly from eq.~(\ref{3.1}).

Suppose, in analogy to what we have just discussed, that the world, as 
known to us, is a movie showing the dynamics of light modes
propagating along two parallel 3-branes
in a five-dimensional space-time, $M$. More precisely, we imagine that 
$M$ is a slab of width $\ell$ in five-dimensional space-time,
${\mathbb R}^5$, the two components, $\partial_+ M$ and $\partial_-
M$, of the boundary of $M$ being identified with the two parallel
3-branes. Let us imagine that, through the five-dimensional bulk $M$
of the system, a massive, charged, four-component spinor field $\psi$
propagates. We consider the response of this system to coupling the
charged fermions described by $\psi$ to a five-dimensional, external
electromagnetic vector potential, $\hat{A}$. By $A_\pm$ we denote
the four-dimensional vector potentials on $\partial_\pm M$ obtained by 
restricting $\hat{A}$ to $\partial_\pm M$. As discussed 
at the end of Sect.~1, there are chiral, left-handed or right-handed,
charged, fermionic surface modes propagating along $\partial_+ M$,
$\partial_-M$, which are coupled to $A_+, A_-$, respectively; see
[6]. In eq.~(\ref{1.74}), the effective action of this system has been 
reported. It is given by 
\begin{eqnarray}
&&{} S_{\rm eff}^E (\hat{A}) \;=\; W_\ell (A_+) + W_r (A_-) -
S_{CS} (\hat{A}) \nonumber\\
&&{}~~~~~~~~~~~+ (4\,\ell\, e^2)^{-1} \int_M d^5 \xi\;
F_{\hat{A}} (\xi)^2 + \ldots~,
\label{3.3}
\end{eqnarray}
where the dots stand for terms $\sim O \left( \frac 1 m \right)$~,
and the renormalization conditions have been chosen in such a way that 
the constant $e^2$ in front of the five-dimensional Maxwell term is
the four-dimensional feinstructure constant. The components,
$\hat{A}_K$, of $\hat{A}$ are denoted by 
\begin{equation}
\hat{A}_\mu \;=:\; A_\mu~,
~~\mu\,=\,0,1,2,3,~~~\hat{A}_4\;=:\;\varphi~, 
\label{3.4}
\end{equation}
i.e., $(\hat{A}_K) = (A,\varphi), ~K = 0,1,2,3,4.$

In order to make contact with the laws of physics in four space-time
dimensions, we should insist on the requirement that left-handed and
right-handed fermions propagating along $\partial_+M$ and $\partial_-
M$, respectively, couple to the {\it same} electromagnetic vector
potential, i.e., that
\begin{equation}
A_+ \left( x, x^4 = \ell\right) \;=\; A_- \left( x, x^4 = 0\right)
\;\equiv\; A(x)~.
\label{3.5}
\end{equation}
This requirement is met if we assume that 
\begin{equation}
\hat{A} \left( x, x^4\right) ~~{\rm is~} independent~{\rm of~} x^4~.
\label{3.6}
\end{equation}
In this case,
\begin{eqnarray}
S_{CS} (\hat{A}) &=& \frac{i\,\ell}{32\,\pi^2} \int_N \varphi
\left( F_A \wedge F_A\right) \nonumber \\
&=& \frac{i\,\ell}{32\,\pi^2} \int_N d^4 x\;\varphi (x) \;
\varepsilon^{\mu\nu\lambda\rho}\, F_{\mu\nu} (x) \, F_{\lambda\rho}
(x) 
\label{3.7}
\end{eqnarray}
where $N \cong {\mathbb R}^4$ is a slice through $M$ parallel to
$\partial_\pm M$, $\mu,\nu =0,1,2,3,$ and $F_A = (F_{\mu\nu})$ is the
four-dimensional field tensor; (the trivial integration over $x^4$ has 
produced the factor $\ell$). Furthermore, the Maxwell term on the
R.S. of (\ref{3.3}) reduces to
\begin{equation}
\frac{1}{4\,e^2}~\left\{ \int_N d^4x\, F_{\mu\nu}(x)\, F^{\mu\nu}
  (x) + \,2 \int_N d^4x\, \left( \partial_\mu \varphi\right) (x)
  \left( \partial^\mu \varphi\right) (x)\right\}~. 
\label{3.8}
\end{equation}
Finally, 
\begin{equation}
W_\ell \left( A_+=A\right) + W_r \left( A_- =A\right) \;=\; S_{\rm
  eff}^E (A)~,
\label{3.9}
\end{equation}
with $S_{\rm eff}^E (A) = S_{\rm eff}^E (A, Z=0)$ as in
eqs.~(\ref{1.29}), (\ref{1.30}). Thus, the complete effective action
of the system is given by
\begin{eqnarray}
S_{\rm eff}^E \left( \varphi; A\right)
&=& S_{\rm eff}^E (A)~+~\frac{i\ell}{32\,\pi^2} \int_N \varphi
\left( F_A \wedge F_A\right) \nonumber\\
&+& \frac{1}{4\,e^2}~\left\{ \int_N d^4x\;F_A^2 (x) +\, 2
  \int_N d^4x \left( \nabla \varphi\right)^2 (x) \right\}~.
\label{3.10}
\end{eqnarray}

Clearly, there is something quite unnatural about this approach: It is 
conditions (\ref{3.5}) and (\ref{3.6})! If $A_+$ were different from
$A_-$ then the fermionic effective action $S_{\rm eff}^E (A) = S_{\rm
  eff}^E (A, Z =0)$ would be replaced by $S_{\rm eff}^E (A,Z)$, where
$A= \frac 1 2 (A_++A_-)$ and $Z = \frac 1 2 (- A_++A_-)$. Thus the
surface modes would not only couple to the electromagnetic field, but
also to a chiral gauge field $Z$ for which there is no experimental
evidence, and the gauge fields would sample a five-dimensional
space-time. 

These unnatural features can be avoided by following {\it Connes'
  formulation} of gauge theories with fermions [16]. Then the
effective action displayed in eq.~(\ref{3.10}) can be reproduced as
follows: One sets $M=N \times {\mathbb Z}_2$, $N \cong {\mathbb R}^4$
and treats the discrete ``fifth dimension'', ${\mathbb Z}_2$, by using
elementary tools from non-commutative geometry [16]. By adding a
``non-commutative'', five-dimensional Chern-Simons action, as
constructed in [17], to Connes' version of the Yang-Mills action (for
a U(1)-gauge field) and to the standard fermionic effective action,
one can reproduce actions like the one in eq.~(\ref{3.10}); see
[17]. There is no room, here, to review the details of these
constructions.

In analogy to what we have discussed above, one may argue that string
theories arise as effective 
theories of surface modes propagating along 9-branes in an
``eleven-dimensional'' space-time, starting from eleven-dimensional
$M$-theory, (with anomalies of the surface theories cancelled by
certain eleven-dimensional Chern-Simons actions). One realization of
this idea appears in [18]. But we shall not pursue these ideas any
further, in this review.

Instead, we ask whether the effective action in (\ref{3.10}) ought to
look familiar to people holding a conventional point of view that
physical space-time is four-dimensional. The answer is ``yes''! The
scalar field $\varphi$ appearing in the effective action on the
R.S. of (\ref{3.10}) can be interpreted as the {\it axion}. The axion
field was originally introduced by Peccei and Quinn [19] to solve the
strong CP problem. There are various reasons, including, primarily,
experimental ones, to feel unhappy about introducing an axion into the 
standard model. But there is also a good reason to do so: String
theory predicts the existence of an axion, the ``model-independent
axion'' first described by Witten [20].

The argument in favor of the model-independent axion goes as follows: 
String theory tells us that there must exist a second-rank
antisymmetric tensor field, i.e., a two-form, $B_{\mu\nu}$. The
gauge-invariant field strength, $H$, a three-form, corresponding to
$B$ is given by
\begin{equation}
H\;=\; dB - \omega_{3YM} + \omega_{3G}~,
\label{3.11}
\end{equation}
where $d$ denotes exterior differentiation, and $\omega_{3YM}$ and
$\omega_{3G}$ are the gauge-field (``Yang-Mills'') and gravitational
(Lorentz) Chern-Simons three-forms. 
(The coefficients in front of these Chern-Simons forms are
proportional to the number, $N_f$, of species of fermions coupled to
the gauge- and gravitational fields. In the following we shall set
$N_f = 1$.)
The field strength $H$ is
invariant under the gauge transformations $B\to B+ d\lambda$, where
$\lambda$ is an arbitrary one-form, and under gauge- and local Lorentz 
transformations accompanied by shifts of $B$. The equation of motion
of $H$ is
\begin{equation}
\partial^\mu\;H_{\mu\nu\lambda} \;=\; 0~,
\label{3.12}
\end{equation}
or $\delta H=0$, where $\delta$ is the co-differential. We consider
the components of $B_{\mu\nu}$ with $\mu,\nu=0,\ldots,3$ and assume
that $B$ is independent of coordinates of {\it internal} dimensions (of
the string theory target). Then, in four-dimensional (non-compact)
space-time, the three-form $H$ is dual to a one-form, $Z$, and the
equation of motion (\ref{3.12}) becomes
\begin{equation}
\partial_\mu\,Z_\nu - \partial_\nu\,Z_\mu \;=\; 0~,~~{\rm
  or~}~d\,Z\;=\;0~. 
\label{3.13}
\end{equation}
By Poincar\'e's lemma, 
\begin{equation}
Z_\mu\;=\; \partial_\mu\;\alpha~,~~{\rm or}~~Z \;=\; d\,\alpha~,
\label{3.14}
\end{equation}
where $\alpha$ is a scalar field. By (\ref{3.11}), the scaling
dimension of $\alpha$ is two. Introducing a constant, $\ell$, with the 
dimension of length, we set
\begin{equation}
\alpha \;=\; \frac{1}{\ell\, e^2} \ \varphi~,
\label{3.15}
\end{equation}
where $\varphi$ has scaling dimension = 1; ($e^2$ is the feinstructure 
constant).

From $d^2=0$ and (\ref{3.11}) we obtain the equation
\begin{equation}
dH(x) \;=\; * {\mathcal A} (x) + {\rm const.}~{\rm tr}\, \left( R (x)
  \wedge R (x)\right)
\label{3.16}
\end{equation}
where
\begin{equation}
* {\mathcal A} (x) \;=\; -~\frac{i}{32\,\pi^2}~ \left( F(x) \wedge
  F(x)\right) 
\label{3.17}
\end{equation}
is the index density, see eq.~(\ref{1.59}), ($*$ denotes the Hodge
dual), and $R (x)$ is the Riemann curvature tensor. Assuming that
space-time is flat, hence $R=0$, and considering the special case,
where the electromagnetic field is the only gauge field in the system, 
we obtain
\begin{equation}
dH \;=\; -~\frac{i}{32\,\pi^2}~\left( F_A \wedge F_A\right)~.
\label{3.18}
\end{equation}
Recalling that 
\[
H \;=\; *~\left( \frac{1}{\ell\,e^2}~d\,\varphi\right)~,
\]
see (\ref{3.13})--(\ref{3.15}), we find that (\ref{3.18}) yields the
following equation of motion for $\varphi$:
\begin{equation}
\square\;\varphi \;=\; -~\frac{i\ell\,e^2}{32\,\pi^2}~\ast\left( F_A
  \wedge F_A\right)~. 
\label{3.19}
\end{equation}
This equation is the Euler-Lagrange equation corresponding to the
action functional
\begin{equation}
\frac{1}{2\,e^2} \int d^4x\,\left(\nabla \varphi\right)^2(x) +
\frac{i\ell}{32\,\pi^2} \int \varphi \left( F_A \wedge F_A\right)~, 
\label{3.20}
\end{equation}
which reproduces the R.S. of (\ref{3.10}), up to the fermionic
effective action and the Maxwell term! The second term in (\ref{3.20}) 
can be understood as arising from coupling {\it fermions} to the
axion. The term in the bare action of the fermions describing their
coupling to the axion is given by
\begin{equation}
\frac{\ell^2}{2} \int d^4x\, H_{\mu\nu\lambda}\,\bar{\psi} \gamma^\mu
  \gamma^\nu \gamma^\lambda \psi \;=\; \frac \ell 2 \int d^4
  x\;\partial_\mu \varphi \bar{\psi} \gamma^\mu \gamma \psi~, 
\label{3.21}
\end{equation}
where $\gamma = \gamma^5$. Carrying out the Berezin integral over the
fermionic degrees of freedom --- see eq.~(\ref{1.29}) --- we find an
effective action for the 
fermions given by
\begin{eqnarray}
S_{\rm eff}^E \left( A, Z \;=\;\frac \ell 2 ~d\varphi\right) 
&=& S_{\rm eff}^E \left( A, Z=0\right)\, -\, i \ell \int
d^4x\;\varphi (x)\;{\mathcal A} (x) \nonumber\\
&=& S_{\rm eff}^E (A)~-~\frac{i\ell}{32\,\pi^2} \int \varphi \left(
  F_A \wedge F_A\right)~, 
\label{3.22}
\end{eqnarray}
in accordance with (\ref{3.20}). The first equation in (\ref{3.22}) is 
eq.~(\ref{1.41}), the second follows from (\ref{1.59}).

Thus, coupling charged Dirac fermions to an external electromagnetic
vector potential $A$ and an axion $\varphi$ yields the effective
action (\ref{3.22}). Adding to it the Maxwell term and the kinetic
energy term for $\varphi$, we again obtain the action (\ref{3.10})!

One may argue that, in any case, the presence of an axion in the
theory may be an 
indication that there must exist {\it extra} (classical or, perhaps
more plausibly, discrete or ``non-commutative'') {\it
  dimensions}. But, for our applications in Sect.~6, this point is not 
important. What {\it will} matter is that the time derivative of the
axion field will play the r\^ole of a, generally speaking, space-time
dependent {\it ``chemical potential''} for right-handed leptons. 

But, quite independently of the properties of fermions (which, for
example, may acquire masses through a Higgs-Kibble mechanism), the
 axion, $\varphi$, will turn out to be the {\it
  driving force} for a possible generation 
of large cosmic magnetic fields.

As our discussion at the beginning of this section, up to
eq.~(\ref{3.10}), has shown it is legitimate to view a
four-dimensional system of fermions in an external electromagnetic and 
an external axion field as the four-dimensional analogue of the edge
degrees of freedom of an incompressible quantum Hall fluid. It
supports electric currents analogous to the diamagnetic edge currents
of a quantum Hall fluid.

\setcounter{equation}{0}
\section{Transport in thermal equilibrium through gapless \\ modes}

In this section we prepare the ground for a theoretical explanation of
effects such as the ones described in Sects.~2 (Examples 1 through 3)
and 3. We consider a quantum-mechanical system ${\mathcal S}$ whose
dynamics is determined by a Hamiltonian $H$, which is a selfadjoint
operator on the Hilbert space ${\mathcal H}$ of pure state vectors of
${\mathcal S}$ with discrete energy spectrum. It is assumed that the
system obeys conservation laws described by some conserved ``charges''
$N_1,\ldots,N_L$  commuting with all observables of the system. Hence
\begin{equation}
\left[ H, N_\ell\right] \;=\; 0~,~~\left[ N_\ell, N_k\right] \;=\;
0~,~~\ell, k \;=\; 1,\ldots,L~,
\label{4.1}
\end{equation}
(e.g. in the sense that the spectral projections of $H$ and of
$N_\ell, N_k$ commute with one another, for all $k$ and $\ell$.) The
system ${\mathcal S}$ is coupled to $L$ reservoirs, ${\mathcal
  R}_1\ldots,{\mathcal R}_L$, with the property that the expectation
value of the conserved charge $N_\ell$ in a stationary state of
${\mathcal S}$ can be tuned to some fixed value through exchange of
``quasi-particles'' between ${\mathcal S}$ and ${\mathcal R}_\ell$,
i.e., through a current between ${\mathcal S}$ and ${\mathcal R}_\ell$ 
that carries ``$N_\ell$-charge'', for all $\ell = 1, \ldots, L$~.

We are interested in describing a thermal equilibrium state of
${\mathcal S}$ coupled to ${\mathcal R}_1,\ldots, {\mathcal R}_L$, at
a temperature $T=(k_B \beta)^{-1}$. According to Gibbs, we should work 
in the grand-canonical ensemble. The reservoirs ${\mathcal R}_1,\ldots
{\mathcal R}_L$ then enter the description of the thermal equilibrium
of ${\mathcal S}$ only through their {\it chemical potentials}
$\mu_1,\ldots,\mu_L$. The chemical potential $\mu_\ell$, is a
thermodynamic  
parameter canonically conjugate to the charge $N_\ell$; in particular, 
the dimension of $\mu_\ell \cdot N_\ell$ is that of an
energy. According to Landau and von Neumann, the thermal equilibrium
state of ${\mathcal S}$ at temperature $(k_B \beta)^{-1}$ in the
grand-canonical ensemble, with fixed values of $\mu_1,\ldots,\mu_L$,
is given by the density matrix
\begin{equation}
\rho_{\beta, \underline{\mu}} \;=\; \Xi_{\beta,
  \underline{\mu}}^{-1}~\exp\, \left[ - \beta \,\left( H -
    \sum_{\ell=1}^L \mu_\ell\,N_\ell\,\right)\right]~, 
\label{4.2}
\end{equation}
where the grand partition function $\Xi_{\beta,\underline{\mu}}$ is
determined by the requirement that
\begin{equation}
{\rm Tr}\;\rho_{\beta,\underline{\mu}} \;=\; 1~.
\label{4.3}
\end{equation}
(It is assumed here that ${\rm exp} \left[ -\beta \left( H - \sum
    \mu_\ell N_\ell\right)\right]$~ is a trace-class operator on
${\mathcal H}$, for all $\beta >0$; we are studying a system in a
compact region of physical space.) The equilibrium expectation of a
bounded operator, $a$, on ${\mathcal H}$ is defined by
\begin{equation}
\langle a \rangle_{\beta,\underline{\mu}} \;:=\; {\rm Tr}~\left(
  \rho_{\beta, \underline{\mu}} \, a\right)~.
\label{4.4}
\end{equation}
Let ${\mathcal J}(x) = \left( {\mathcal J}^0 (x), \underline{{\mathcal J}}
(x)\right)$ be a conserved quantum-mechanical current density of
${\mathcal S}$, where $x  
= (\underline{x},t)$, $t$~is time and $\underline{x}$ is a point of
physical space contained inside ${\mathcal S}$. We are interested in
calculating the expectation values of products of components of
${\mathcal J}$ in the state $\rho_{\beta, \underline{\mu}}$; in
particular, we should like to calculate $\langle \underline{{\mathcal J}}
(x)\rangle_{\beta, \underline{\mu}}$. Of course, if the dimension of
space is larger than one, $\langle \underline{{\mathcal J}}
(x)\rangle_{\beta, \underline{\mu}}$ vanishes unless rotation
invariance is broken by some external field. If $\underline{{\mathcal
    J}}(x)$ is a vector current then $\langle \underline{{\mathcal J}}
(x)\rangle_{\beta, \underline{\mu}}$ vanishes unless the state
$\rho_{\beta,\underline{\mu}}$ is {\it not} invariant under
space-reflection and time reversal. This happens if some of the
charges $N_1,\ldots,N_L$ are not invariant under space-reflection and
time reversal, i.e., if they are {\it chiral}.

To say that ${\mathcal J}$ is conserved means that it satisfies the
continuity equation
\begin{equation}
\partial_\mu\,{\mathcal J}^\mu \;=\; 0~,
\label{4.5}
\end{equation}
where $x^0=t$ denotes time, and $\partial_\mu = \partial/\partial
x^\mu$. If the space-time of the system ${\mathcal S}$ is
topologically trivial (``star-shaped'') then eq.~(\ref{4.5}) implies
that there is a globally defined vector field $\underline{\varphi}(x)$ such
that 
\begin{equation}
{\mathcal J}^0(x) \;=\; \frac{q}{2\pi}\; {\rm div}\, \underline{\varphi}
(x)~,~~\underline{{\mathcal J}}  (x) \;=\; -\,\frac{q}{2\pi}
~\frac{\partial}{\partial 
  t}~\underline{\varphi} (x)~, 
\label{4.6}
\end{equation}
with $q$ the electric charge.

Let us suppose that $\underline{\varphi}(x)$ is an operator-valued
distribution on ${\mathcal H}$, whose time-dependence is determined by 
the formal Heisenberg equation
\begin{equation}
\frac{\partial}{\partial t} \ \underline{\varphi} (x) \;=\; \frac i \hbar
~\left[ H, \underline{\varphi} (x) \right]~.
\label{4.7}
\end{equation}
[Technically, we are treading on somewhat slippery ground here; but we 
shall proceed formally, in order to explain the key ideas on a few
pages.] From (\ref{4.6}) and (\ref{4.7}) we derive that
\begin{equation}
\langle \underline{{\mathcal J}} (x)\rangle_{\beta, \underline{\mu}} \;=\;
\frac{iq}{h}~ \langle \left[ H, \underline{\varphi} (x)
\right]\rangle_{\beta,\underline{\mu}}~. 
\label{4.8}
\end{equation}
Formally, the R.S. of (\ref{4.8}) {\it vanishes}, because $\langle
(\cdot)\rangle_{\beta,\underline{\mu}}$ is a time-translation
invariant state. However, the field $\underline{\varphi}$ turns out to have
ill-defined {\it zero-modes}, and it is not legitimate to pretend that 
$[H,\underline{\varphi}(x)] = H \underline{\varphi}(x) -
\underline{\varphi}(x) H$, 
because both terms on the R.S. are divergent, due to the zero-modes of 
$\underline{\varphi}$. What {\it is} legitimate is to claim that 
\begin{equation}
\frac{\partial}{\partial t}~\underline{\varphi} (x) \;=\; \frac i
h~\left[ H - \sum_{\ell=1}^L \mu_\ell N_\ell, \underline{\varphi}
  (x)\right] \;+\; \frac i h \sum_{\ell=1}^L \mu_\ell \left[
  N_\ell, \underline{\varphi} (x) \right]~,
\label{4.9}
\end{equation}
and that the expectation value
\[
\bigg\langle \left[ H - \sum_{\ell=1}^L \mu_\ell N_\ell,
  \underline{\varphi} 
  (x) \right] \bigg\rangle_{\beta,\underline{\mu}}
\]
{\it vanishes}. This can be seen by replacing the Hamiltonian $H$ by a 
{\it regularized} Hamiltonian $H^{(\varepsilon)}$ generating a
dynamics that eliminates the zero-modes of $\underline{\varphi}$. One
replaces the state $\rho_{\beta,\underline{\mu}}$ by a regularized
state $\rho_{\beta,\underline{\mu}}^{(\varepsilon)}$ proportional to
$\exp \left[ -\beta \left( H^{(\varepsilon)} - \sum \mu_\ell N_\ell
  \right)\right]$, and we set
\[
\langle a\rangle_{\beta,\underline{\mu}}^{(\varepsilon)} \;:=\; {\rm
  tr}\,\left( \rho_{\beta, \underline{\mu}}^{(\varepsilon)} a\right)~, 
\]
for any bounded operator $a$ on ${\mathcal H}$. Then
\begin{eqnarray}
\langle \underline{{\mathcal J}} (x)\rangle_{\beta,\underline{\mu}} 
&=& \lim_{\varepsilon \to 0}~ \frac{iq}{h}~\bigg\langle \left[
  H^{(\varepsilon)}\;- \sum_{\ell=1}^L \mu_\ell N_\ell, \underline{\varphi}
  (x)\right] \bigg\rangle_{\beta,\underline{\mu}}^{(\varepsilon)}
\nonumber\\ 
&+& \lim_{\varepsilon\to 0}~ \sum_{\ell=1}^L \frac{i q
  \mu_\ell}{h}~\langle \left[ N_\ell, \underline{\varphi} (x)
\right]\rangle_{\beta, \underline{\mu}}^{(\varepsilon)}~.
\label{4.10}
\end{eqnarray}
Obviously
\begin{equation}
\bigg\langle \left[ H^{(\varepsilon)}\;-\; \sum_{\ell=1}^L \mu_\ell
  N_\ell, \underline{\varphi} (x)\right]
\bigg\rangle_{\beta,\underline{\mu}}^{(\varepsilon)} \;=\; 0~,
\label{4.11}
\end{equation}
and one might be tempted to expect that
$\displaystyle\mathop{\lim}_{\varepsilon\to 0} 
\langle \left[ N_\ell, \underline{\varphi} (x) \right]
\rangle_{\beta,\underline{\mu}}^{(\varepsilon)}$ vanishes, for all
$\ell$, because the charges $N_\ell$ are conserved. However, as long
as the regularization is present $(\varepsilon \neq 0)$, these charges 
are {\it not} conserved, and there is no guarantee that the second
term on the R.S. of (\ref{4.10}) vanishes!

We conclude that 
\begin{eqnarray}
\langle \underline{{\mathcal J}} (x)\rangle_{\beta,\underline{\mu}} 
&=&
\lim_{\varepsilon\to 0}~ \sum_{\ell=1}^L~\frac{iq\mu_\ell}{h}~
\langle \left[ N_\ell, \underline{\varphi} (x)\right]\rangle_{\beta,
  \underline{\mu}}^{(\varepsilon)} \nonumber\\
&=:& \sum_{\ell=1}^L~ \frac{iq\mu_\ell}{h}~ \langle \left[ N_\ell, 
  \underline{\varphi}(x) \right] \rangle_{\beta,\underline{\mu}} ~.
\label{4.12}
\end{eqnarray}
Eq.~(\ref{4.12}) might be called a {\it current sum rule}.

Let us assume that the conserved charges $N_\ell, \ell = 1,2,\ldots, $ 
are given as integrals of the 0-components of conserved currents over
space. Then the current sum rule (\ref{4.12}) implies that if $\langle 
\underline{{\mathcal J}} (x) \rangle _{\beta, \underline{\mu}} \neq 0$
there  
must be {\it gapless modes} in the system. The proof,  see [7], is
analogous to the proof of the {\it Goldstone theorem} in the theory of 
broken continuous symmetries. 

The sum rule (\ref{4.12}) is the main result of this section. A
careful derivation of equation (\ref{4.12}) and of our analogue of the 
Goldstone theorem could be given by using the {\it operator-algebra
  approach} to quantum statistical mechanics [21]. But, in order to
reach our punch line on a reasonable number of pages, we refrain from
entering into a careful technical discussion.

\setcounter{equation}{0}
\section{Conductance quantization in ballistic wires and in
  incompressible quantum Hall fluids}

In this section, we combine the results of Sects.~2 and 4, in order to 
gain insight into the phenomena of conductance quantization, as
discussed at the beginning of Sect.~2. We first study a ballistic
wire, i.e., a very thin, long, clean conductor without back scattering 
centers (impurities). The ends of the wire are connected to two
reservoirs filled with electrons at chemical potentials $\mu_\ell,
\mu_r$, respectively, with
\begin{equation}
\mu_\ell - \mu_r \;=\; V~,
\label{5.1}
\end{equation}
where $V$ is the voltage drop through the wire. 

A ballistic wire is a three-dimensional, 
elongated metallic object with a tiny cross section in the plane
perpendicular to its principal axis. Thus, at low temperature, the
three-dimensional nature of the wire merely implies that there are
several, say $N$, species of electrons labelled by discrete quantum
numbers that originate from the motion in the plane perpendicular to
the axis of the wire. Every species of electrons forms a {\it
  one-dimensional Luttinger liquid} [22], and these Luttinger liquids
may interact with each other. Every Luttinger liquid has {\it two}
conserved vector current operators, ${\mathcal J}^{(i,s) \mu}$, and
conserved {\it chiral} current operators, $\widehat{{\mathcal
    J}}_{\ell/r}^{(i,s)\mu}$, where $s = \uparrow, \downarrow$ denotes 
the magnetic quantum number of the electrons in the i$^{\rm th}$ 
Luttinger liquid (``spin up'' and ``spin down''), and
$i=1,\ldots,N$. The chiral current operators $\widehat{{\mathcal
    J}}_\ell^{(i,s)\mu}$ are as in
eqs.~(\ref{1.21})--(\ref{1.23}). The total electric current operator
and the total chiral current operators are given by
\begin{equation}
{\mathcal J}^\mu
\;=\;\sum_{i=1 \atop s=\uparrow,\downarrow}^N
{\mathcal J}^{(i,s)\mu}~,~~ \widehat{{\mathcal J}}_{\ell/r}^\mu \;=\;
\sum_{i=1\atop s=\uparrow,\downarrow}^N
\widehat{{\mathcal J}}_{\ell,r}^{(i,s)\mu}~.
\label{5.2}
\end{equation}
They are conserved. The total electric charge operators counting the
electric charges of chiral (left-moving and right-moving) modes in the 
wire are the operators $N_\ell$ and $N_r$ defined in
eq.~(\ref{1.24}). Their expectation values in a thermal equilibrium
state of the wire are tuned by the chemical potentials, $\mu_\ell,
\mu_r$, respectively, of the reservoirs at the right and left end of
the wire.

Imagine that the wire is kept at a constant temperature
$\beta^{-1}$. Our description of the electron gas in the wire in terms 
of a finite number of Luttinger liquids correctly captures electric
transport properties of the wire {\it only} if $\beta^{-1}$ and
$eV$, with $e$ the elementary electric charge, are {\it tiny}
as compared to the energy scale of the motion in the plane
perpendicular to the axis of the wire. (However, $\beta^{-1}$ and $eV$ 
should be {\it large} as compared to the energy scale of weak back
scattering centers.) We shall assume that these conditions are
met. Then we may apply the current sum rule (\ref{4.12})
derived in the last section, and the formulae for the anomalous
commutators derived in Sect.~1, see (\ref{1.16}) and the equation
after (\ref{1.24}), in order to calculate the electric current, $I$,
in the wire corresponding to a voltage drop $V$. The current sum rule
(\ref{4.12}) yields
\begin{eqnarray}
I &=& \langle \underline{{\mathcal J}}
(x)\rangle_{\beta,\underline{\mu}} \nonumber\\
&=& \frac{i q}{h}~\left\{ \mu_\ell\,\big\langle \left[ N_\ell,
    \underline{\varphi} (x)\right]\big\rangle_{\beta,\underline{\mu}} + 
  \mu_r\, \big\langle \left[ N_r,
    \underline{\varphi}(x)\right]\big\rangle_{\beta,\underline{\mu}}
\right\}~,  
\label{5.3}
\end{eqnarray}
where $\underline{\varphi}$ is the potential of the current ${\mathcal 
  J}^\mu$. Since the currents ${\mathcal J}^{(i,s)\mu}$ of all the
Luttinger liquids are conserved, every one of them can be derived from 
a potential, $\varphi^{(i,s)}$, 
\begin{equation}
{\mathcal J}^{(i,s)\mu} (x) \;=\; \frac{q}{2\pi}~\varepsilon^{\mu\nu}
\left( \partial_\nu\,\varphi^{(i,s)}\right) (x)~,
\label{5.4}
\end{equation}
see eq.~(\ref{1.11}), and $q=-e$, because the electric charge of an
electron is equal to minus the elementary electric charge.

Plugging (\ref{5.4}) and (\ref{5.2}) into eq.~(\ref{5.3}) and
recalling eq.~(\ref{1.24}) and the anomalous commutator
\begin{eqnarray}
&&{} \left[ \widehat{{\mathcal J}}_{\ell/r}^{(i,s)0}
  \left(\underline{y},t\right),~\varphi^{(i',s')} \left(
    \underline{x},t\right)\right] \nonumber\\
&&~~~~~=~\pm i~\frac{e}{2\pi}~\delta_{ii'}\, \delta_{ss'}\,\delta
\left( \underline{x} - \underline{y}\right)~, 
\label{5.5}
\end{eqnarray}
see eqs.~(\ref{1.11}), (\ref{1.15}), (\ref{1.16}), we find that
\begin{eqnarray}
I &=& -~\frac{ie}{h}~\sum_{i=1 \atop s=\uparrow,\downarrow}^N \left\{
  \mu_\ell \bigg\langle \left[ N_\ell^{(i,s)},~\varphi^{(i,s)} \left(
      \underline{x},
      t\right)\right]\bigg\rangle_{\beta,\underline{\mu}}
\right.\nonumber\\ 
&&{}~~~~~~~~~~~~~~~~~~\left. +~\mu_r\,\bigg\langle \left[
  N_r^{(i,s)},~\varphi^{(i,s)}\,
  \left(\underline{x},t\right)\right]\bigg\rangle_{\beta,
  \underline{\mu}}\right\}  \nonumber\\
&=& \frac{e^2}{h} \times 2N \times \left(\mu_\ell - \mu_r\right)
\nonumber\\ 
&=& 2N~\frac{e^2}{h}~V~.
\label{5.6}
\end{eqnarray}
Thus, we have derived the formula
\begin{equation}
G_W \;=\; \frac I V \;=\; 2N~\frac{e^2}{h}~,
\label{5.7}
\end{equation}
as claimed in Example 2 at the beginning of Sect.~2.

Of course, the number, $N$, of Luttinger liquids of electrons in the
wire depends on the mean {\it Fermi energy} of the wire (at zero
temperature) and hence on the electron density in the wire and can be
tuned.

The quantization of the {\it Hall conductance} of an incompressible
Hall fluid in a Hall sample with e.g. an annular (Corbino) geometry
(see Example~1) can be understood by using very similar arguments as
in the example of quantum wires. Let $V$ denote 
the voltage drop between the outer and the inner edge of the
sample. We assume that $eV$ and the temperature $\beta^{-1}$ are tiny, 
as compared to the mobility gap in the bulk of the fluid. Let us also
assume, {\it temporarily}, that the electric field created by
connecting the outer and inner edge to the two leads of a battery with 
voltage drop $V$ does {\it not} penetrate into the bulk of the sample
(i.e., that, in the bulk, it is screened completely). If this
assumption (which will actually turn out to be irrelevant, later) is
made then the entire Hall current, $I_H$, in the sample is carried by
the chiral modes propagating along
the edges of the sample, i.e., $I_H$ is given by the expectation value 
of the sum, $\widehat{{\mathcal J}}_\ell^1 + \widehat{{\mathcal
    J}}_r^1$, of the edge currents, $\widehat{{\mathcal J}}_\ell^\mu,
  \widehat{{\mathcal J}}_r^\mu$. For an appropriate choice of
  orientation, $\widehat{{\mathcal J}}_\ell^\mu$ is the current at the 
  outer edge and $\widehat{{\mathcal J}}_r^\mu$ is the current at the
  inner edge of the sample. The two edges are separated by the bulk,
  and, for a {\it macroscopic} sample, tunnelling of quasi-particles
  from one edge to the other one can be neglected for all practical
  purposes. This implies that the currents, $\widehat{{\mathcal
      J}}_\ell^\mu$ and $\widehat{{\mathcal J}}_r^\mu$, and hence the
  charge operators $N_\ell$ and $N_r$ defined in eq.~(\ref{1.24}), are 
  conserved to very high accuracy. The anomalous commutators of
  $\widehat{{\mathcal J}}_\ell^\mu$ and $\widehat{{\mathcal J}}_r^\mu$ 
  are given in eq.~(\ref{2.15}), and the analogue of Eqs.~(\ref{1.11}) and 
  (\ref{5.4}) is
\begin{equation}
{\mathcal J}^\mu (x) \;=\;
e~\frac{\sqrt{f_H}}{2\pi}~\varepsilon^{\mu\nu}~\left( \partial_\nu
  \varphi\right) (x)~.
\label{5.8}
\end{equation} 
Inserting these equations into the current sum rule (\ref{4.12}), one
finds that
\begin{eqnarray}
I_H &=& -\;\frac e h ~\sqrt{f_H}~\left\{ \mu_\ell \big\langle \left[
    N_\ell, \varphi \left(
      \underline{x},t\right)\right]\big\rangle_{\beta,\underline{\mu}} 
  + \mu_r \big\langle \left[ N_r, \varphi \left(
      \underline{x},t\right)\right]\big\rangle_{\beta,\underline{\mu}}
  \right\} \nonumber\\ 
&=& \frac{e^2}{h}~f_H\, \left( \mu_\ell - \mu_r\right) \nonumber\\
&=& \sigma_H~V~.
\label{5.9}
\end{eqnarray}
These arguments do not make it clear why the Hall fraction $f_H =
\left( e^2/h\right)^{-1} \sigma_H$ is a {\it rational number}, and we
have no clue, so far, which rational numbers may turn up in physical
samples. Understanding the rational quantization of $f_H$ is not quite 
an easy matter; see
[8, 11]. Here we can only sketch some key ideas. Let $\psi
(\underline{x}, t)$ denote the field (a ``chiral vertex operator'')
creating an electron or a hole propagating along the inner (or along 
the outer) edge of the sample. This field has the form
\begin{equation}
\psi \left( \underline{x},t\right) \;=\; :\,
e^{iq\varphi_{\ell/r}(\underline{x},t)}\;:\; \varepsilon \left(
  \underline{x},t\right)~, 
\label{5.10}
\end{equation}
where $q$ is a real number to be determined, $\varphi_{\ell/r}
(\underline{x},t)$ is the potential of the conserved chiral edge
current, i.e., it is a massless, chiral free field, and $\varepsilon
(\underline{x},t)$ is an electrically neutral so-called simple current 
of a {\it rational} chiral conformal field theory describing chiral
modes of zero charge propagating along the edge. The field $\psi
(\underline{x},t)$ must carry electric charge 
$\pm e$. Using formula (\ref{5.8}) and recalling that $\varepsilon$
has zero electric charge, we find that
\begin{equation}
q \;=\; 1 \big/ \sqrt{f_H} ~.
\label{5.11}
\end{equation}
Furthermore, the field $\psi (\underline{x},t)$ must obey Fermi
statistics (because electrons and holes are fermions). Hence it must
have half-integer ``conformal spin'', i.e.,
\begin{equation}
s_\psi \;\equiv\; 1/2~{\rm mod.~}1~.
\label{5.12}
\end{equation}
By eq.~(\ref{5.10}), the conformal spin of $\psi$ is given by
\begin{equation}
s_\psi \;=\; \frac{q^2}{2}~+~s_\varepsilon \;=\; \frac{1}{2\,f_H}~+~
s_\varepsilon~, 
\label{5.13}
\end{equation}
where $s_\varepsilon$ is the conformal spin of $\varepsilon$. Because
$\varepsilon$ is a simple current of a {\it rational} chiral conformal 
field theory, 
$s_\varepsilon$ is a {\it rational number}, i.e., $s_\varepsilon =
\frac k \ell$, with $k$ and $\ell$ two relatively prime integers. Thus
(\ref{5.12}) and (\ref{5.13}) imply that 
\begin{equation}
\frac{1}{2\,f_H}~+~\frac k \ell ~\equiv~ 1/2~{\rm mod.~} 1~.
\label{5.14}
\end{equation}
It follows that $f_H$ is a {\it rational number}. For more details see 
[8, 23, 24] and, especially, [11]. Properties of the rational chiral
conformal  
field theories that may appear in the context of the quantum Hall
effect are discussed in [8, 11]. One noteworthy result is that, unless 
$f_H$ is an {\it integer}, there must be chiral modes
(quasi-particles) of fractional electric charge and fractional
statistics, sometimes called Laughlin vortices, propagating along the
edges of the sample.

Let us see what happens if the electric field $\underline{E}$ {\it
  can} penetrate into the bulk of an incompressible quantum Hall
fluid. Electric transport in such Hall fluids can be understood by
combining the arguments outlined above with Hall's law in the
bulk. The total Hall current, $I_H$, is given by
\begin{equation}
I_H \;=\; I_H^{\rm edge} + I_H^{\rm bulk}~,
\label{5.15}
\end{equation}
where $I_H^{\rm edge}$ is the edge current studied above, and
$I_H^{\rm bulk}$ is a current carried by extended bulk states. Let
$\gamma$ denote an arbitrary smooth oriented curve connecting a point
on the inner edge to a point on the outer edge of the
sample. Then
\begin{equation}
I_H^{\rm bulk} \;=\; - \sum_{k,\ell} \int_\gamma J^k \left(
  \underline{x}, t\right)\;\varepsilon_{k\ell}\;ds^\ell \left(
  \underline{x}\right)~, 
\label{5.16}
\end{equation}
where $J^k$ is the $k$-component of the bulk current; see
eq.~(\ref{2.4}). As usual, 
\begin{equation}
J^k \left( \underline{x},t\right) \;=\; \big\langle {\mathcal J}^k \left(
  \underline{x},t\right)\big\rangle_A \;=\; \delta\,S_{\rm eff} (A)
\big/ \delta \,A_k \left( \underline{x},t\right) ~.
\label{5.17}
\end{equation}
By eqs.~(\ref{2.17}), (\ref{2.18}), the R.S. of (\ref{5.17}) is given
by 
\begin{equation}
\delta\,S_{\rm eff} (A) \big/ \delta\,A_k \left( \underline{x},
  t\right) \;=\; \sigma_H\;\varepsilon_{k\ell}\;E^\ell \left(
  \underline{x},t\right)~, 
\label{5.18}
\end{equation}
see also (\ref{2.4}) (with $\sigma_L=0$). Thus
\begin{equation}
I_H^{\rm bulk} \;=\; \sigma_H \int_\gamma \underline{E} \left(
  \underline{x},t\right)\,\cdot\, d\underline{s} (\underline{x})~.
\label{5.19}
\end{equation}
We have shown in eq.~(\ref{5.9}) that
\begin{equation}
I_H^{\rm edge} \;=\; \sigma_H\, \left( \mu_\ell - \mu_r\right)~.
\label{5.20}
\end{equation}
 Thus, combining (\ref{5.15}), (\ref{5.19}) and (\ref{5.20}), we
 conclude that
\begin{equation}
I_H \;=\; I_H^{\rm edge} + I_H^{\rm bulk} \;=\; \sigma_H\,\left(
  \mu_\ell - \mu_r + \int_\gamma \underline{E} \left(
    \underline{x},t\right) \,\cdot\, d \underline{s} \left(
    \underline{x}\right)\right)~. 
\label{5.21}
\end{equation}
But the expression in the parenthesis on the R.S. of (\ref{5.21}) is
nothing but the {\it total voltage drop} $V$ between the outer and the 
inner edge. Hence (\ref{5.21}) implies that
\begin{equation}
I_H \;=\; \sigma_H\;V~,
\label{5.22}
\end{equation}
as desired.

Transport phenomena such as {\it heat conduction} through a quantum
wire or a Hall sample (see Example~3 at the beginning of Sect.~2) can
be studied along similar lines: In a physical system where modes of
{\it different} chirality do not interact with each other (such as the 
modes at the inner and at the outer edge of the sample containing an
incompressible Quantum Hall fluid) the left-moving and the right
moving modes can be coupled to different reservoirs at {\it
  different} temperatures $\beta_\ell^{-1}$ and $\beta_r^{-1}$. This
results in a {\it non-zero} heat current given by an expectation value 
of the component $T^{01}$ of the energy-momentum tensor of the
conformal field theory describing the chiral modes in an equilibrium
state where the left-movers are at temperature $\beta_\ell^{-1}$ and
the right-movers at temperature $\beta_r^{-1} \neq
\beta_\ell^{-1}$. (Such expectation values can be calculated from
Virasoro characters.) These ideas lead to a conceptually clean
understanding of the effects described in Example~3 at the beginning
of Sect.~2. 

\setcounter{equation}{0}
\section{A four-dimensional analogue of the Hall effect, and the
  generation of large, cosmic magnetic fields in the early universe}

In this section, we further explore the four-dimensional analogue of
the Hall effect described in Sect.~3. We shall apply our findings to
exhibit effects that may play an important r\^ole in early-universe
cosmology. Our results represent an elaboration upon those in [15, 7].

We start our analysis by studying a system of massless Dirac fermions
coupled to an external electromagnetic field in four-dimensional
Minkowski space. Using results derived in Sects.~1 and 4, we derive
equations analogous to eqs.~(\ref{5.3})--(\ref{5.6}) for the
conductance of a quantum wire.

From Sect.~1 we recall the expression for the {\it anomalous
  commutators} between vector- and axial-vector --- or chiral
currents.
\begin{equation}
\left[ \widehat{{\mathcal J}}_{\ell/r}^0 \left( t,
    \underline{x}\right),~\widehat{{\mathcal J}}_{\ell/r}^0 \left( t,
    \underline{y}\right) \right] \;=\;
\pm~i~\frac{q^2}{4\pi^2}~\left( \underline{B} \left( \underline{x},
    t\right) \,\cdot\, \underline{\nabla}\right)\; \delta \left(
  \underline{x}-\underline{y}\right)~,  
\label{6.1}
\end{equation}
where $q$ is the charge of the fermions --- see eq.~(\ref{1.62}) ---
and
\begin{equation}
\left[ \widehat{{\mathcal J}}_\ell^0 \left( t, \underline{x}\right)\,, 
  ~\widehat{{\mathcal J}}_r^0 \left( t, \underline{y}\right)\right]
  \;=\; 0~.
\label{6.2}
\end{equation}
With (\ref{1.45}) and (\ref{1.48}), these equations yield 
\begin{equation}
\left[ \widehat{{\mathcal J}}_{\ell/r}^0 \left( t,
    \underline{y}\right)\,,~{\mathcal J}^0 \left( t,
    \underline{x}\right)\right] \;=\; \pm~i~\frac{q^2}{8\pi^2}~
\left( \underline{B} \left( \underline{y}, t\right)\,\cdot\,
  \underline{\nabla}_{\underline{x}}\right) ~\delta \left(
  \underline{x} - \underline{y}\right)~,
\label{6.3}
\end{equation}
where ${\mathcal J}^\mu$ is the $\mu$-component of the conserved
vector current. In Sect.~4, we have introduced the vector potential,
$\underline{\varphi}$, of ${\mathcal J}^\mu$:
\begin{equation}
{\mathcal J}^0 (x) \;=\; \frac{q}{2\pi}~{\rm div}\;
\underline{\varphi} (x)\,,~ \underline{{\mathcal J}} (x) \;=\; -\,
\frac{q}{2\pi}~\frac{\partial}{\partial t}~ \underline{\varphi} (x)~.
\label{6.4}
\end{equation}
Eqs.~(\ref{6.3}) and (\ref{6.4}) imply that
\begin{eqnarray}
\left[ \widehat{{\mathcal J}}_{\ell/r}^0 \left( \underline{y},
    t\right)\,,~\underline{\varphi} \left( \underline{x},
    t\right)\right] 
&=& \pm~i~\frac{q}{4\pi}~\underline{B} \left( \underline{y},
  t\right)~\delta \left( \underline{x} - \underline{y}\right)
\nonumber\\
&\pm& {\rm curl~} \underline{\Pi} \left( \underline{x} -
  \underline{y}, t \right)
\label{6.5}
\end{eqnarray}
 where $\underline{\Pi}$ is some vector-valued distribution.

Next, we recall that the operators 
\begin{equation}
N_{\ell/r} \; :=\; \int d
\underline{y}~\widehat{{\mathcal J}}_{\ell/r}^0 \; \left(
  \underline{y}, t\right)
\label{6.6}
\end{equation}
are {\it conserved}. They are interpreted as the electric charge
operators for left-handed/right-handed fermionic modes. The chemical
potentials conjugate to $N_{\ell/r}$ are denoted by
$\mu_{\ell/r}$. Let us imagine that, at {\it very early times} in the
evolution of our universe (or others), there was an asymmetry in the
population of left-handed and right-handed fermionic modes, (as argued 
in [15] for the example of electrons before the electroweak phase
transition). Then
\begin{equation}
\mu_\ell~\neq~\mu_r~,
\label{6.7}
\end{equation}
in the state of the universe at those very early times. Let us
furthermore imagine that the state of the universe at those early
times was, to a good approximation, a thermal equilibrium state at an
inverse temperature $\beta$ ( $\lesssim \left( 80~ TeV\right)^{-1}$, as
argued in [15]) and with chemical potentials $\mu_\ell$ and
$\mu_r$. (It may well be that this is an unrealistic assumption. ---
It will subsequently turn out that it is unimportant!)

Under these assumptions, we may apply the {\it current sum rule}
(\ref{4.12}) derived in Sect.~4. Combining eqs.~(\ref{6.5}),
(\ref{6.6}) and (\ref{4.12}), and using that $\int\limits_{{\mathbb R}^3}
d\underline{y}~{\rm curl~} \underline{\Pi} \left( \underline{x} -
  \underline{y}, t\right) = 0$, for all $\underline{x} , t$, we find
that
\begin{eqnarray}
\big\langle \underline{{\mathcal J}} (x)\big\rangle_{\beta,
  \underline{\mu}} &=& \frac{iq}{h}~\left\{ \mu_\ell \big\langle
  \left[ N_\ell, \underline{\varphi} (x) \right]
  \big\rangle_{\beta,\underline{\mu}} \;+\; \mu_r \;\big\langle \left[ 
    N_r, \underline{\varphi} (x)\right]\big\rangle_{\beta,
    \underline{\mu}}\right\} \nonumber\\
&=& -~\frac{q^2}{4\pi \,h}~\left( \mu_\ell -
  \mu_r\right)~\underline{B} (x)~,
\label{6.8}
\end{eqnarray}
as claimed in [7]. This equation is the analogue of (\ref{5.6}).

Treating the electromagnetic field as a classical, but {\it dynamical} 
field, its dynamics is governed by Maxwell's equations, 
\[
\underline{\nabla}\,\cdot\,\underline{B} \;=\;0~,~~\underline{\nabla}
\wedge \underline{E} \;+\; \partial_t\, \underline{B} \;=\; 0~,
\]
and
\begin{equation}
\underline{\nabla}\,\cdot\,\underline{E} \;=\; \langle {\mathcal
  J}^0\rangle_{\beta, \underline{\mu}}~, ~~\underline{\nabla} \wedge
\underline{B} - \partial_t \, \underline{E} \;=\; \langle
\underline{{\mathcal J}}\rangle_{\beta, \underline{\mu}}~. 
\label{6.9}
\end{equation}
There is no reason to imagine that the charge density, $\langle
{\mathcal J}^0\rangle_{\beta, \underline{\mu}}$, in the very early
universe is different from zero. In the last equation of (\ref{6.9}),
the current on the R.S. is given by eq.~(\ref{6.8}). Actually,
assuming that there are some {\it dissipative processes} evolving in
the early universe, an equation for the current, 
\[
\underline{J} (x) \;:=\; \langle \underline{{\mathcal J}}
(x)\rangle_{\beta, \underline{\mu}}~,
\]
more realistic than (\ref{6.8}) may be 
\begin{equation}
\underline{J} (x) \;=\; \sigma_L \, \underline{E}\, (x) \;+\;
\sigma_T\, V \, \underline{B}\, (x)~,
\label{6.10}
\end{equation}
where $\sigma_L$ is an Ohmic longitudinal conductivity, and
\begin{equation}
\sigma_T \;:=\; -~\frac{q^2}{4\pi\,h}
\label{6.11}
\end{equation}
is the analogue of the {\it ``transverse''} or {\it Hall
  conductivity}; furthermore,
\begin{equation}
V \;:=\; \mu_\ell - \mu_r
\label{6.12}
\end{equation}
is the analogue of the {\it voltage drop} considered in the Hall
effect. The quantity $\sigma_T$ is {\it ``quantized''}, just like the
Hall conductivity: If there are $N>1$ species of charged, massless
fermions, with electric charges $q_1,\ldots,q_N,$ then
\begin{equation}
\sigma_T \;=\; -~\frac{1}{4\pi\,h}~ \left( \sum_{j=1}^N q_j^2\right)~,
\label{6.13}
\end{equation}
which is the precise analogue of a formula for the quantization of the 
Hall conductivity derived in [8], and, for $q_j = \pm\, e,
~j=1,\ldots,N,$ of eq.~(\ref{5.6}).

Let us temporarily assume that $\sigma_L=0$, (i.e., we 
neglect dissipative processes). Then Maxwell's equations, together
with eq.~(\ref{6.10}) (for $\sigma_L=0$) and the assumption that the
charge density vanishes, yield the following system of linear
equations:
\begin{eqnarray}
&&{} \underline{\nabla}\,\cdot\,\underline{B}\;=\;0~,~~
\underline{\nabla} \wedge \underline{E} \;+\;
\partial_t\,\underline{B} \;=\;0~,\nonumber\\
&&{} \underline{\nabla}\,\cdot\,\underline{E}\;=\;0~,~~
\underline{\nabla}\wedge\underline{B}\,-\,\partial_t \,\underline{E}
\;=\; \sigma_T\,V\,\underline{B}~.
\label{6.14}
\end{eqnarray}
Because all coefficients are constant, these equations can be solved
by Fourier transformation, and it is enough to construct propagating
wave solutions corresponding to an arbitrary, but fixed wave vector
$\underline{k}$. The equations $\underline{\nabla} \cdot \underline{B} 
= \underline{\nabla} \cdot\underline{E} =0$ imply that
\begin{equation}
\underline{k}\,\cdot\,\widehat{\underline{B}}\;=\;
\underline{k}\,\cdot\,\widehat{\underline{E}}\;=\;0~, 
\label{6.15}
\end{equation}
i.e., that only the components of the Fourier transforms
$\widehat{\underline{B}}$ and $\widehat{\underline{E}}$ of
$\underline{B}$ and $\underline{E}$ (evaluated at the wave vector
$\underline{k}$) perpendicular to $\underline{k}$ can be
non-zero. Denoting the components of $\widehat{\underline{B}}$ and
$\widehat{\underline{E}}$ 
perpendicular to $\underline{k}$ by
$\widehat{\underline{B}}^T,~\widehat{\underline{E}}^T$, respectively,
the remaining equations in (\ref{6.14}) yield
\begin{equation}
\partial_t~{~\widehat{\underline{E}}^T \choose
  ~\widehat{\underline{B}}^T~} ~=~ K(\underline{k})~
{~\widehat{\underline{E}}^T \choose ~\widehat{\underline{B}}^T~}~,
\label{6.16}
\end{equation}
where (in an orthonormal basis chosen in the plane perpendicular to
$\underline{k}$) the matrix $K (\underline{k})$ is given by
\begin{equation}
K\,(\underline{k})\;=\;
\left( \begin{array}{cccc}
~0~    &~0~    &~-\,\sigma_T\,V~ &~-\,i\,k~\\
~0~    &~0~    &i\,k~            &~-\,\sigma_T\,V~\\
~0~    &~i\,k~ &~0~              &~0~\\
~-\,i\,k~ &~0~    &~0~              &~0~
\end{array}\right)
\label{6.17}
\end{equation}
with $k=|\underline{k}|$. The circular frequency of a propagating wave 
solution of (\ref{6.14}) with wave vector $\underline{k}$ is given by
$\omega (\underline{k})$, where $i\,\omega (\underline{k})$ is an
eigenvalue of $K (\underline{k})$. By (\ref{6.17}),
\begin{equation}
\omega\,(\underline{k})^2 \;=\; k^2~\pm~k\,\sigma_T\,V~,
\label{6.18}
\end{equation}
as one readily checks. Thus, if
\begin{equation}
|\underline{k}| \;=\; k \,<\, \sigma_T\,V
\label{6.19}
\end{equation}
there are two purely imaginary frequencies, and eqs.~(\ref{6.14}) have 
solutions $\left( \underline{B} \left(
    \underline{x},t\right),~\underline{E}
  \left(\underline{x},t\right)\right)$ growing {\it exponentially fast 
  in time} and with the property that
\begin{equation}
\underline{B} \left( \underline{x},t\right)\,\cdot\,\underline{E}
\left( \underline{x},t\right) \;\neq\; 0~.
\label{6.20}
\end{equation}
It is almost as easy to solve Maxwell's equations (\ref{6.9}), with
$\underline{J}$ given by (\ref{6.10}), for $\sigma_L \neq 0$. For wave 
vectors $\underline{k}$ satisfying
\begin{equation}
\sigma_L^2\;<\; |\underline{k}|\;\sigma_T\,V \;<\; \left(
  \sigma_T\,V\right)^2~, 
\label{6.21}
\end{equation}
one again finds exponentially growing electromagnetic fields;
(perturbation theory). Dissipative processes will subsequently damp
out electric fields.

In [15], calculations similar to those just presented are used to
argue that, in the very early universe, large, cosmic electromagnetic
fields may have been generated as a consequence of an asymmetric
population of left-handed and right-handed electron modes $( q = -
e)$. However, these
arguments rest on rather shaky hypotheses; (the state of the early
universe is assumed to be a thermal equilibrium state, and the charges 
$N_\ell$ and $N_r$, see eq.~(\ref{6.6}), are assumed to be
approximately conserved). We propose to reconsider these arguments in
the light of the analogy between the (2+1)-dimensional (bulk)
description of the Hall effect and the (4+1)-dimensional description
of chiral fermions discussed at the beginning of Sect.~3,
eqs~(\ref{3.3}) through (\ref{3.10}). What we have described, so far,
in this section are calculations analogous to those reported in
eqs.~(\ref{5.6}), (\ref{5.8}) and (\ref{5.9}). Next, we generalize our 
analysis in a way analogous to that followed in eqs.~(\ref{5.15})
through (\ref{5.22}), starting from the effective action given in
(\ref{3.10}); (see also (\ref{3.20})).

We integrate out all degrees of freedom (quarks, gluons, leptons, the
weak gauge fields --- $W, Z$ --- etc.), except for the {\it
  electromagnetic} and the {\it axion field}. We have seen, at the
beginning of Sect.~3, eqs.~(\ref{3.4}), (\ref{3.10}), that the axion
could be viewed as the four-component of a five-dimensional
electromagnetic vector potential, $\widehat{A}$, which does not depend 
on the coordinate, $x^4$, in the direction perpendicular to the
four-dimensional branes on which we live; see (\ref{3.6}). We could
pursue a five- (or higher-) dimensional approach to early-universe
cosmology (as presently popular), --- but let's not! We propose to
view the axion as the ``model-independent (invisible)'' axion first
described in [20]. It has a geometrical origin (in superstring
theory). It couples to {\it all} gauge fields present in the system
through a term
\begin{equation}
\frac{i\,\ell}{32\,\pi^2} \int \varphi \left( F_W \wedge F_W\right)~,
\label{6.22}
\end{equation}
where $F_W$ is the field strength of a gauge field $W$ appearing in
our theoretical description, and to the curvature tensor $R$; see
(\ref{3.16}). All gauge fields, except for the electromagnetic vector
potential $A$, shall be integrated out. The (Euclidian-region-)
functional integrals have the form
\begin{equation}
\int d\mu (W)~{\rm exp~} \left[ \frac{i\,\ell}{32\,\pi^2} \int \varphi 
  \left( F_W \wedge F_W\right)\right] \;=:\; e^{-\,U(\varphi)}~.
\label{6.23}
\end{equation}
Since $\frac{i}{32\,\pi^2}~\left( F_W \wedge F_W\right)$ is the index
density, the integrand in $U (\varphi)$ can be shown to be {\it periodic}
in $\varphi$, for $\varphi$ independent of $x$, with period
$\frac{2\pi}{\ell}~$. It is known that (somewhat loosely speaking)
$d\mu$ is a positive measure and that it is invariant under space
reflection, which changes the sign of $\int F_W \wedge F_W$. It
follows that~ ${\rm exp} \left( - U (\varphi)\right)$ is real and has
its maxima at $\varphi = \frac{2\pi}{\ell}~n,~~n=0, \pm 1, \pm 2,
\ldots~.$
(See e.g. [25] for more details.)

A transition amplitude from a configuration $\left( A_{\rm in},
  \varphi_{\rm in}\right)$ of the electromagnetic --- and the axion
field at a very early time, $t_1$, to a configuration $\left( A_{\rm
    out}, \varphi_{\rm out}\right)$ at a much later time, $t_2$, can
be computed from the Feynman path integral
\begin{equation}
\int {\mathcal D} A\;{\mathcal D} \varphi~ e^{i\,S_{\rm eff} (A,
  \varphi) / \hbar}~,
\label{6.24}
\end{equation}
with boundary conditions $\left( A(t_1), \varphi (t_1)\right) = \left( 
  A_{\rm in}, \varphi_{\rm in}\right)$ and $\left( A (t_2), \varphi
  (t_2)\right) = \left( A_{\rm out}, \varphi_{\rm out}\right)$. In
(\ref{6.24}), $S_{\rm eff} (A, \varphi)$ denotes the {\it total}
effective action over Minkowski space. It is obtained from $S_{\rm
  eff}^E (A, \varphi)$, the effective action in the Euclidian region,
by undoing the Wick rotation described in eq.~(\ref{1.28}). By
eqs.~(\ref{3.10}) or (\ref{3.20}) and (\ref{6.23}), $S_{\rm eff} (A,
\varphi)$ has the general form
\begin{eqnarray}
S_{\rm eff} (A, \varphi)
&=& \frac{1}{4\,e^2} \int d^4x \left\{ F_{\mu\nu} (x)\, F^{\mu\nu}
  (x)\,+\, 2\,\left( \partial_\mu \varphi\right) (x) \left(
    \partial^\mu \varphi\right) (x)\right\} \nonumber\\
&+& \frac{\ell}{32\,\pi^2} \int \varphi (x) \left( F \wedge F\right)
(x) \,-\, U (\varphi) \,+\, W(A)~, 
\label{6.25}
\end{eqnarray}
where $W(A)$ is of higher than second order in $A$ and arises from
integrating out all charged fields in the
theory\renewcommand{\thefootnote}{\fnsymbol{footnote}}\footnote[1]{$W$ 
  depends on the boundary conditions, at times $t_1, t_2$, imposed on
  the fields that have been integrated
  out.}\renewcommand{\thefootnote}{\arabic{footnote}};
furthermore, $e^2$ is the effective (one-loop renormalized)
feinstructure constant. It is {\it not} necessary, in this approach,
to assume that all the fermions in the theory be massless. They can
acquire masses through the Higgs--Kibble mechanism. (The arguments of
complex chiral Higgs fields then contain a term proportional to the
axion field $\varphi$ which, however, can be absorbed in a change of
variables.) Furthermore, calculating transition amplitudes with the
help of Feynman path integrals does not presuppose that the system is
in or close to thermal equilibrium. 

We now insert expression (\ref{6.25}) into the functional integral
(\ref{6.24}) and try to evaluate the latter by using a semi-classical
expansion based on the stationary-phase method. The equations for the
saddle point are
\begin{equation}
\delta\,S_{\rm eff} (A, \varphi) \big/ \delta A_\mu (x)\;=\;0~,
~~\delta\,S_{\rm eff} (A, \varphi) \big/ \delta\varphi (x) \;=\;0~.
\label{6.26}
\end{equation}
To simplify matters, we consider solutions of these equations
describing fairly {\it small} electromagnetic fields and an axion
field that varies only {\it slowly} in space-time. Then we can neglect 
the term $W(A)$ in (\ref{6.25}) and we may omit all contributions to
$U(\varphi)$ involving {\it derivatives}, $\partial_\mu\varphi$, of
the axion field $\varphi$. The saddle point equations (\ref{6.26})
then yield the following coupled Maxwell--Dirac-axion equations: 
\begin{eqnarray}
\partial_\mu\,F^{\mu\nu}
&=& \frac{\ell\,e^2}{8\,\pi^2}~\partial_\mu\,\left( \varphi\,
  \widetilde{F}^{\mu\nu}\right)~,\nonumber\\ 
\square \;\varphi
&=& \frac{\ell\,e^2}{32\,\pi^2}~*~\left( F \wedge F\right) - U'
(\varphi)~, 
\label{6.27}
\end{eqnarray}
(and we have set $c=1$ and $\hbar =1$). Let $J_M^\mu$ denote the
magnetic current that could be present if there were magnetic
monopoles moving through the early universe. Then the full set of
Maxwell--Dirac-axion equations reads
\begin{eqnarray}
\partial_\mu\,\widetilde{F}^{\mu\nu}
&=& J_M^\nu~, ~~\partial_\mu\,F^{\mu\nu}\;=\;
\frac{\ell\,e^2}{8\,\pi^2}~ \left\{ \left( \partial_\mu
    \varphi\right)\; \widetilde{F}^{\mu\nu}\,+\, \varphi\,J_M^\nu
\right\}~,\nonumber\\
\square\;\varphi
&=& \frac{\ell\,e^2}{32\,\pi^2}~*~\left( F \wedge F\right) \,-\,
U'(\varphi)~. 
\label{6.28}
\end{eqnarray}
The first equation in (\ref{6.28}) replaces the homogeneous Maxwell
equations, \\ $\left( \partial_\mu \widetilde{F}^{\mu\nu} = 0,~~ {\rm for~}
  J_M^\nu = 0\right)~.$ In vector notation, the system of equations
(\ref{6.28}) reads
\begin{eqnarray}
\underline{\nabla}\,\cdot\, \underline{B}
&=& J_M^0~, ~~\underline{\nabla} \wedge \underline{E}\;+\;
\dot{\underline{B}}\;=\; \underline{J}_M~,\nonumber\\
\underline{\nabla}\,\cdot\,\underline{E}
&=& \frac{\ell\,e^2}{8\,\pi^2}~\left\{ \left(
    \underline{\nabla}\,\varphi\right)\,\cdot\,
  \underline{B}\;+\;\varphi\,J_M^0\right\}~, \nonumber\\
\nabla \wedge \underline{B}
&-& \dot{\underline{E}}\;=\;-\;\frac{\ell\,e^2}{8\,\pi^2}~ \left\{
  \dot{\varphi}\, \underline{B} \;+\; \underline{\nabla}\, \varphi \wedge
  \underline{E} \;+\; \varphi\,\underline{J}_M \right\}~, \nonumber\\
\square\;\varphi
&=& -\;\frac{\ell\,e^2}{8\,\pi^2}~\underline{E} \,\cdot\,\underline{B}
\;-\; U' (\varphi)~.
\label{6.29}
\end{eqnarray}
In order to gain some insight into properties of solutions of these
highly {\it non-linear} equations, we study their linearization around
various special solutions. Already this part of the analysis, let
alone a study of the full, non-linear equations, is quite lengthy; see
[26] for a beginning. Here we just sketch results in a few interesting
special situations.

We shall first assume that $J_M^\mu \equiv 0$, i.e., that there aren't
any magnetic monopoles around.

\medskip

(i)~~ We set $U(\varphi) = 0$ and consider the following special
solution of eqs.~(\ref{6.29}). 
\begin{eqnarray}
&&{} \underline{E}\;=\; \underline{B}\;\equiv\;0~, \nonumber\\
&&{} \varphi \left( \underline{x},t\right) \;=\; \frac V \ell~\cdot\; t~,
\label{6.30}
\end{eqnarray}
where $V$ is a constant. Linearizing (\ref{6.29}) around (\ref{6.30}), 
we obtain the equations
\begin{eqnarray}
\underline{\nabla}\,\cdot\,\underline{B} &=& 0~, ~~\underline{\nabla}
\wedge \underline{E} \;+\; \dot{\underline{B}}\;=\;0~, \nonumber\\
\underline{\nabla}\,\cdot\,\underline{E} &=& 0~, ~~\underline{\nabla}
\wedge \underline{B}\;-\;
\dot{\underline{E}}\;=\;-\;\frac{e^2}{8\,\pi^2}~ V\, \underline{B}~,
\nonumber\\
\square\;\varphi &=& 0~.
\label{6.31}
\end{eqnarray}
With the exception of the wave equation for the axion field $\varphi$, 
these equations are {\it identical} to eqs.~(\ref{6.14}), with
$\sigma_T = \frac{e^2}{8\,\pi^2}~.$ Had we not set $\hbar = 1$, the
equation for $\sigma_T$ would read
\[
\sigma_T \;=\; -\;\frac{e^2}{4\,\pi\, h}~,
\]
which is precisely eq.~(\ref{6.11}), with $q=e$! Recall that, in the
analysis presented at the beginning of this section,
\[
V \;=\; \mu_\ell\;-\;\mu_r~.
\]
This equation and (\ref{6.30}) tell us that, apparently, the field
$\ell \dot{\varphi}$ has the interpretation of the {\it difference of
  chemical potentials} of {\it left-} and {\it right-handed fermions}!
This interpretation magically fits with the {\it five-dimensional} 
interpretation of the axion field $\varphi$ as the four-component,
$\widehat{A}_4$, of an electromagnetic vector potential $\widehat{A}$
defined over a slab of height $\ell$ in five-dimensional Minkowski
space; see eqs.~(\ref{3.4}), (\ref{3.6}) and (\ref{3.10}). Then
\[
\dot{\varphi} \;=\; \dot{\widehat{A}}_4 \;=\; E_4
\] 
is the four-component of the electric field. Integrating $E$ along an
oriented curve, $\gamma$, joining a point on the lower face of the
slab to a point on the upper face, at fixed time, we obtain
\begin{equation}
\int\limits_\gamma \sum_{K=1}^4 E_K \,\cdot\,ds^K \;=\;
\int\limits_0^\ell dx^4\, E_4 (\xi) \;=\; \int\limits_0^\ell dx^4\,
\dot{\varphi} (\xi)~,
\label{6.32}
\end{equation}
where $\xi = (x,x^4) = (t, \underline{x}, x^4)$, and we have assumed
in the first equality that $E$ does not depend on $x^4$ (see
assumption (\ref{3.6})) and $E_4$ does not depend on
$\underline{x}$. Since, for solution (\ref{6.30}), 
\[
E_4 (\xi) \;=\; \dot{\varphi} (\xi) \;=\; \frac V \ell ~,
\]
eq.~(\ref{6.32}) yields
\begin{equation}
\int\limits_\gamma \sum_{K=1}^4 E_K \,\cdot\, ds^K \;=\; V~.
\label{6.33}
\end{equation}
This shows that, in the five-dimensional interpretation of the axion,
$V$ is the ``voltage drop'' between the two four-dimensional branes
corresponding to the lower and upper face of the five-dimensional
slab. This observation makes the analogy between the effects studied
here and the Hall effect yet a little more precise.

Solutions of eqs.~(\ref{6.31}) have been studied earlier in this
section; see (\ref{6.16}) through (\ref{6.20}).
They have unstable modes growing exponentially in time, with
$\underline{B} (\underline{x},t) \,\cdot\,\underline{E}
(\underline{x},t) \neq 0$~. 

\medskip

(ii)~~ Now $U (\varphi) \neq 0$; $U' (\varphi) (x) := \delta U
(\varphi)\big/ \delta \varphi (x)$ is a periodic function with
minima at $\frac{2\pi}{\ell}~n$, $n = 0, \pm 1, \pm 2, \ldots$~. We
linearize equations (\ref{6.29}) around the solution
$\underline{E} = \underline{B} \equiv 0~$, $\varphi = \varphi_c (t)$~, 
where $\varphi_c (t)$ solves the equation
\begin{equation}
\ddot{\varphi} (t) \;=\;-\;U'\left( \varphi \left(t\right)\right)~.
\label{6.34}
\end{equation}
This is the equation of motion of a planar pendulum in a force field
with potential $U$. We have learnt in our courses on elementary
mechanics how to solve (\ref{6.34}), using energy conservation. For
``small energy'', a solution, $\varphi_c(t)$, of (\ref{6.34}) is a
periodic function of $t$; for ``large energy'', $\varphi_c(t)$ grows
linearly in $t$, with periodic modulations superimposed; and
$\dot{\varphi}_c (t)$ is periodic in $t$. 

Eqs.~(\ref{6.29}), with $J_M^\mu \equiv 0$, linearized around
$\underline{E} = \underline{B} = 0~,$ $\varphi = \varphi_c (t)~,$
yield the equations
\begin{eqnarray}
&&{} \underline{\nabla}\,\cdot\,\underline{B} \;=\;0~,
~~\underline{\nabla} \wedge \underline{E} \;+\; \dot{\underline{B}}
\;=\; 0~,\nonumber\\
&&{} \underline{\nabla}\,\cdot\,\underline{E}\;=\;0~,
~~\underline{\nabla} \wedge \underline{B}\,-\,\dot{\underline{E}}
\;=\;-\;\frac{\ell\,e^2}{8\,\pi^2} ~\dot{\varphi}_c\,\underline{B}~, 
\label{6.35}
\end{eqnarray}
which can be solved by Fourier transformation in the space
variables. The equations for the components,
$\widehat{\underline{B}}^T$ and $\widehat{\underline{E}}^T$, of the
Fourier components of $\underline{B}$ and $\underline{E}$
perpendicular to the wave vector $\underline{k}$ are two Mathieu
equations of the form

\[
{~\dot{\xi}~ \choose ~\dot{\eta}~} ~=~ 
\left( \begin{array}{ll}
0 &1\\
h_k(t)~~ &0 
\end{array}\right)
~ {~\xi~ \choose ~\eta~}~,
\]
where $k = |\underline{k}|$ and $h_k (t)$ depends on $k$ and is linear 
in $\dot{\varphi}_c (t)$; see [26]. These equations yield
\begin{equation}
\ddot{\xi} (t) \;=\; h_k (t)\;\xi(t)~.
\label{6.36}
\end{equation}
In solving this equation one encounters the phenomenon of the {\it
  parametric resonance}, i.e., for $k$ in a family of intervals,
eq.~(\ref{6.36}) has a solution growing {\it exponentially} in
time. Hence the electromagnetic field has {\it unstable modes} growing
exponentially in time and with $\underline{B} \cdot \underline{E} \neq 
0$. 

\medskip

The parametric resonance has appeared in cosmology in other
contexts. In our analysis it plays an entirely natural and essentially 
{\it model-independent} r\^ole and may help to explain where large,
cosmic (electro) magnetic fields might come from. 

Of course, eqs.~(\ref{6.29}) are Lagrangian equations of motion. They 
are derived from the action functional (\ref{6.25}), (with $W=0$ and
$U$ independent of derivatives of $\varphi$). The Lagrangian density
does not depend on time explicitly. Therefore, there is a {\it
  conserved energy functional}, ${\mathcal E} (A, \varphi)$. The
special solutions considered in (\ref{6.30}) and (\ref{6.34}) have
{\it infinite} (axionic) energy. The instabilities in the time
evolution of the electromagnetic field are due to a reshuffling of
energy from axionic to electromagnetic degrees of freedom. 

Clearly, it would be interesting to construct finite-energy solutions
of eqs.~(\ref{6.29}), with an initial axion field depending not only
on time but also on space. Of particular significance is situation
(ii), with $U\neq 0$. Interpreting $\ell \dot{\varphi}$ as a
difference of chemical potentials for left- and right-handed fermions, 
we are thus considering states of the universe with spatially varying, 
time-dependent chemical potentials triggering an asymmetric population 
of left-handed and right-handed fermionic modes. This asymmetry
gradually disappears, due to chirality-changing processes, and the
field energy stored in axionic degrees of freedom is reshuffled into
certain electromagnetic field 
modes triggering the growth of cosmic electromagnetic fields. Large
electric fields rapidly die out because of dissipative processes; (the 
energy loss from the electric field into matter degrees of freedom is
described by $\underline{E} \cdot \underline{J} \propto \sigma_L
|\underline{E}|^2 + \ldots~.$) But large magnetic fields may survive
for a comparatively long time.

Describing these phenomena within the approximation of linearizing
eqs.~(\ref{6.29}) (possibly supplemented by a dissipative Ohmic term)
around special solutions, including space-dependent ones, of infinite
or finite energy, is feasible; [26]. But our understanding of the
effects of the {\it non-linearities} in eqs.~(\ref{6.29}) remains, not 
surprisingly, very rudimentary.

Some speculations on the r\^ole played by magnetic monopoles in the
effects described here are
contained in the last section; see also [26].

\setcounter{equation}{0}
\section{Conclusions and outlook}

In this review we have shown how the chiral, abelian anomaly helps to
explain important features of the (quantum) {\it Hall effect}, such as 
the existence of edge currents and aspects of the quantization of the
Hall conductivity, and of its {\it four-dimensional cousin}, which may 
play a significant r\^ole in explaining the origin of large, cosmic
magnetic fields. Our analysis is essentially {\it model-independent},
a fact that makes it quite trustworthy. How significant the
four-dimensional variant of the Hall effect is in early-universe
cosmology remains to be understood in more detail. This will require
a better understanding of orders of magnitude of various physical
quantities and of the properties of solutions of the non-linear
Maxwell--Dirac-axion equations (\ref{6.29}). A beginning has been made 
in [15, 26]. --- There is no doubt that the following equations 
\begin{equation}
J_{\rm bulk} \;=\; \sigma_T \,*\,F~, ~~\delta\, J_{\rm edge} \;=\; 
-\;\sigma_T\,E~, 
\label{7.1}
\end{equation}
with $\sigma_T = \sigma_H$, for bulk- and edge-currents of an
incompressible Hall fluid (see eqs.~(\ref{2.10}) and (\ref{2.14})),
and
\begin{equation}
J^\nu \;=\; \sigma_T\,\ell\,\left\{ \left(
    \partial_\mu\,\varphi\right)\; \widetilde{F}^{\mu\nu} \;+\;
  \varphi\, J_M^\nu\,\right\}~,
\label{7.2}
\end{equation}
where $\sigma_T = - \frac{1}{4\,\pi\,h}~\left( \sum_{j=1}^N
  q_j^2\right)~,$ with $N$ the number of species of charged fermions
with electric charges $q_1, \ldots, q_N$, (see eqs.~(\ref{6.13}) and
(\ref{6.28})) are significant laws of nature connected with the chiral
anomaly.

For the future, it would be important to gain a better understanding
of the contents of equations (\ref{6.29}), (possibly corrected by
dissipative terms and/or ones coming from $\delta W (A)\big / \delta
A_\mu (x)$, which have been neglected), including the r\^ole played by 
magnetic monopoles and dyons $\left( J_M^\mu \neq
  0\right)$. (Eqs.~(\ref{6.29}) and their fully quantized counterparts 
appear to offer some clue for understanding (axion-driven)
monopole--anti-monopole annihilation, triggering the growth of certain 
modes of the electromagnetic field.) Some understanding of these
issues has been gained in [26]; but much work remains to be done. We
have also studied  the influence of gravitational fields on the
processes described in Sect.~6 [26] (in analogy to the ``geometric'' (or
gravitational) Hall effect in 2+1 dimensions described in the third
paper quoted under [10] and to the phenomenon of ``quantized'' heat
currents in quantum wires mentioned in Sects.~3 and 5). But there is
no room here to describe our results in detail. Our
findings will have to be combined with cosmic evolution equations. 


\centerline{------------------------}

\medskip

In this review, we have only quoted literature that we used in
carrying out the calculations described here. Many further references may be
found in [7, 8, 10, 15, 20, 26].

\section*{Acknowledgments} The results described in Sects.~2, 4
and 5 have been obtained in collaboration (of J.F.) with A. Alekseev
and V. Cheianov [7], in continuation of earlier work with T. Kerler,
U. Studer and E. Thiran. We thank these colleagues, Chr.~Schweigert
and Ph.~Werner for many useful discussions. We are grateful to
R. Durrer, E. Seiler and D. Wyler for drawing our attention to some
useful earlier work in the literature and for encouragement.

\end{document}